\documentclass[12pt,authoryear, 3p]{elsarticle}

\usepackage{amssymb}
\usepackage{amsmath}
\usepackage{amsthm}

\journal{Computational Statistics \& Data Analysis}

\usepackage[plain]{fancyref}
\usepackage{mathtools}
\usepackage{physics}
\usepackage[x11names, dvipsnames]{xcolor}
\usepackage{graphicx}
\usepackage{enumerate}
\usepackage{makecell}
\usepackage{booktabs}
\usepackage{pdflscape}
\usepackage{afterpage}
\RequirePackage[colorlinks,citecolor=blue,linkcolor=blue,urlcolor=blue]{hyperref}

\theoremstyle{plain}
\newtheorem{theorem}{Theorem}
\newtheorem{corollary}[theorem]{Corollary}

\newtheorem{innerassumption}{Assumption}
\newenvironment{assumption}[1]
{\renewcommand\theinnerassumption{#1}\innerassumption}
{\endinnerassumption}
\theoremstyle{remark}

\newcommand{\R}{\mathbb{R}}  

\newcommand{\N}{\mathbb{N}}
\newcommand{\Q}{\mathbb{Q}}

\renewcommand{\P}{\mathbb{P}}

\newcommand{\F}{\mathcal{F}}

\newcommand{\vt}{\vartheta}
\newcommand{\vtn}{\hat{\vt}_n}
\newcommand{\eps}{\varepsilon}
\newcommand{\indset}[1]{{I_{ #1 }}}
\newcommand{\ind}[1]{{\indset{\left\{ #1 \right\}}}}
\newcommand{\expec}[2][]{\mathbb{E}_{#1}\left[#2\right]}
\newcommand{\cov}[2][]{\text{COV}_{#1}\left(#2\right)}

\newcommand{\prob}[2][{}]{\mathbb{P}_{#1}\left(#2\right)}

\newcommand{\betan}{\hat{\beta}_n}

\newcommand{\del}{\partial}

\newcommand{\opI}{o_{\mathbb{P}}(1)}
\newcommand{\sumin}{\sum_{i=1}^n}
\newcommand{\sumjk}{\sum_{j=1}^k}

\newcommand{\cspace}{C[-\infty, \infty]}

\newcommand{\lspace}[1][\bar{\R}]{\ell^\infty(#1)}

\newcommand{\proc}[2]{\alpha_{#1}^{(#2)}}
\newcommand{\nproc}[1]{\proc{n}{#1}}

\newcommand{\tproc}[1]{\tilde{\alpha}_{#1} }
\newcommand{\tnproc}{\tproc{n}}
\newcommand{\tinfproc}{\tproc{\infty}}
\newcommand{\tsproc}[1]{\tilde{\alpha}_{#1}^{*} }
\newcommand{\tsnproc}{\tsproc{n}}

\newcommand{\Fn}{F_{Y,n}}

\newcommand{\Fo}{F_{Y,0}}

\newcommand{\Ys}{Y_{i,n}^*}
\newcommand{\Xs}{X_{i,n}^*}
\newcommand{\vtns}{\vtn^*}
\newcommand{\probs}[1]{\mathbb{P}_n^*\qty(#1)}
\newcommand{\expecs}[1]{\mathbb{E}_n^*\left[#1\right]}
\newcommand{\covs}[1]{\text{COV}_n^*\left(#1\right)}
\newcommand{\vars}[1]{\text{Var}_n^*\left(#1\right)}
\newcommand{\opIs}{o_{\mathbb{P}_n^*}(1)}

\newcommand{\fancyrefequlabelprefix}{eq}
\frefformat{plain}{\fancyrefequlabelprefix}{equation\fancyrefdefaultspacing(#1)}
\Frefformat{plain}{\fancyrefequlabelprefix}{Equation\fancyrefdefaultspacing(#1)}

\newcommand{\fancyreftheoremlabelprefix}{thm}
\frefformat{plain}{\fancyreftheoremlabelprefix}{Theorem\fancyrefdefaultspacing#1}
\Frefformat{plain}{\fancyreftheoremlabelprefix}{Theorem\fancyrefdefaultspacing#1}

\newcommand{\fancyrefcorollarylabelprefix}{cor}
\frefformat{plain}{\fancyrefcorollarylabelprefix}{Corollary\fancyrefdefaultspacing#1}
\Frefformat{plain}{\fancyrefcorollarylabelprefix}{Corollary\fancyrefdefaultspacing#1}

\newcommand{\fancyrefdefinitionlabelprefix}{def}
\frefformat{plain}{\fancyrefdefinitionlabelprefix}{Definition\fancyrefdefaultspacing#1}
\Frefformat{plain}{\fancyrefdefinitionlabelprefix}{Definition\fancyrefdefaultspacing#1}

\newcommand{\fancyreflemmalabelprefix}{lem}
\frefformat{plain}{\fancyreflemmalabelprefix}{Lemma\fancyrefdefaultspacing#1}
\Frefformat{plain}{\fancyreflemmalabelprefix}{Lemma\fancyrefdefaultspacing#1}

\newcommand{\fancyrefpropositionlabelprefix}{prop}
\frefformat{plain}{\fancyrefpropositionlabelprefix}{Proposition\fancyrefdefaultspacing#1}
\Frefformat{plain}{\fancyrefpropositionlabelprefix}{Proposition\fancyrefdefaultspacing#1}

\newcommand{\fancyrefremarklabelprefix}{rem}
\frefformat{plain}{\fancyrefremarklabelprefix}{Remark\fancyrefdefaultspacing#1}
\Frefformat{plain}{\fancyrefremarklabelprefix}{Remark\fancyrefdefaultspacing#1}

\newcommand{\fancyrefexamplelabelprefix}{ex}
\frefformat{plain}{\fancyrefexamplelabelprefix}{Example\fancyrefdefaultspacing#1}
\Frefformat{plain}{\fancyrefexamplelabelprefix}{Example\fancyrefdefaultspacing#1}

\newcommand{\fancyrefassumptionlabelprefix}{ass}
\frefformat{plain}{\fancyrefassumptionlabelprefix}{Assumption\fancyrefdefaultspacing#1}
\Frefformat{plain}{\fancyrefassumptionlabelprefix}{Assumption\fancyrefdefaultspacing#1}

\newcommand{\fancyrefconditionlabelprefix}{cond}
\frefformat{plain}{\fancyrefconditionlabelprefix}{(#1)}
\Frefformat{plain}{\fancyrefconditionlabelprefix}{(#1)}

\begin{document}

\begin{frontmatter}

\title{Bootstrap-Based Goodness-of-Fit Test for Parametric Families of Conditional Distributions} %

\author[inst1]{Gitte Kremling\corref{cor1}} %
\author[inst1]{Gerhard Dikta}

\cortext[cor1]{Corresponding Author. Email: kremling@fh-aachen.de}

\affiliation[inst1]{organization={FH Aachen University of Applied Sciences, Department of Medical Engineering and Technomathematics},%
            addressline={Heinrich-Mußmann-Str.\ 1}, 
            city={Jülich},
            postcode={52428}, 
            country={Germany}}

\begin{abstract}
	A consistent goodness-of-fit test for distributional regression is introduced. The test statistic is based on a process that traces the difference between a nonparametric and a semi-parametric estimate of the marginal distribution function of $Y$. As its asymptotic null distribution is not distribution-free, a parametric bootstrap method is used to determine critical values. 
	Empirical results suggest that, in certain scenarios, the test outperforms existing specification tests by achieving a higher power and thereby offering greater sensitivity to deviations from the assumed parametric distribution family. Notably, the proposed test does not involve any hyperparameters and can easily be applied to individual datasets using the gofreg-package in R.
\end{abstract}

\begin{keyword}

Model checking \sep Specification test \sep Distributional regression \sep Monte Carlo simulation

\end{keyword}

\end{frontmatter}

\makeatletter
\def\ps@pprintTitle{%
\let\@oddhead\@empty
\let\@evenhead\@empty
\def\@oddfoot{\centerline{Accepted for publication in Computational Statistics \& Data Analysis: \href{https://doi.org/10.1016/j.csda.2025.108289}{doi:10.1016/j.csda.2025.108289}}}%
\let\@evenfoot\@oddfoot}
\makeatother

\section{Introduction}
\label{sec:intro}

In many scientific applications, a response variable together with a number of features that may influence the outcome is observed. It is of high interest to figure out in which way the response (random) variable $Y \in \R$ depends on the vector of input (random) variables $X \in \R^p$. In this paper, we will propose a test to check whether the conditional distribution of $Y$ given $X$ fits into a given parametric family. According to \citet{andrews1997conditional}, many models in micro-econometric and biometric applications are of this type, and \citet{maddala1983limited} and \citet{mccullagh1983generalized} list numerous examples. 
A common class of parametric regression models used in practice is the generalized linear model (GLM). It was first introduced in \citet{nelder1972generalized} and later on thoroughly discussed in \citet{fahrmeir2013regression, fox2018r, dikta2021bootstrap}.

The test problem we want to consider is defined as follows. Let $\{(X_i, Y_i)\}_{i=1}^n$ be an i.i.d.\ sample of covariates and response variables with distribution function $H$ for $X$ and conditional density function $f(\,\cdot\,|x)$ for $Y$ given $X=x$ with respect to a $\sigma$-finite measure $\nu$.
We want to create a test for
\begin{align}
	\label{eq:gof_hyp}
	H_0: f \in \mathcal{F} = \{(x,y) \mapsto f(y|\vt,x) ~|~ \vt \in \Theta\} \quad\text{vs.}\quad H_1: f \notin \mathcal{F},
\end{align}
where 
$$ \Theta \coloneqq \left\{\vt \,\Big|\, \int \int f(y|\vt,x) \nu(dy) H(dx) < \infty \right\} \subseteq \R^q $$
defines the set of admissible parameters. If $H_0$ holds, we denote the true distribution parameter by $\vt_0$. 
Note that, as opposed to classical regression, which merely assumes a model for the conditional mean $m(x) = \expec{Y|X=x}$, we consider a model for the whole distribution of $Y$ given $X$. This enables us not only to predict the value of $Y$ for a new feature vector $X$ but also, for example, to provide a confidence band for the estimate. \citet{kneib2023rage} offers a thorough investigation of distributional regression models, advocating for their superiority compared to mean regression and discussing the four currently most prominent model classes.

The literature on goodness-of-fit tests for conditional distributions and related model checks for parametric families of regression functions is very extensive. The methods can be categorized into two general approaches, namely those that make use of nonparametric kernel estimators and those that do not. Representatives of the former class are, for example, given in \citet{rodriguez1998testing, zheng2000consistent, fan2006nonparametric, pardo2007goodness, cao2008goodness, ducharme2012omnibus}. Since methods of this type suffer from the problem of choosing an appropriate smoothing parameter, we suggest a model check that falls into the second category and compare it to other representatives of this class.

In \cite{andrews1997conditional}, a so-called conditional Kolmogorov (CK) test is proposed for the test problem defined in \fref{eq:gof_hyp}. It is based on a process tracing the difference between a nonparametric and semi-parametric estimate of the joint distribution of $X$ and $Y$ defined by
\begin{equation}
	\label{eq:proc_andrews}
	\nu_n(t, x) = \frac{1}{\sqrt{n}} \sumin \qty(\ind{Y_i \le t} - F\big(t|\vtn, X_i\big)) \ind{X_i \le x}, \quad (t,x) \in \R \times \R^p,
\end{equation}
where $\vtn$ denotes an estimator for the distribution parameter $\vt$. The corresponding CK test statistic is given by $\sup_{j = 1, \dots, n} \abs{\nu_n(Y_j, X_j)}$. A practical problem of this test may occur for high-dimensional covariate vectors as, even for large sample sizes, the inequality $X_i \le X_j$ may never be satisfied for $i \ne j$. In this case, the test statistic collapses to $$\sup_{j = 1, \dots, n} \frac{1}{\sqrt{n}}(1-F(Y_j|\vtn, X_j)) \le \frac{1}{\sqrt{n}},$$
which means that, regardless of the true underlying conditional distribution, the test asymptotically never rejects the null hypothesis! An example application with $X \in \R^{16}$ where this issue is encountered can be found in \citet[\mbox{Sec.\ 4}]{bierens2012integrated}.

Another approach, first introduced in \cite{stute2002model} and later on extended by a bootstrap method in \cite{dikta2021bootstrap}, relies on the estimated marked empirical process in estimated direction, which is given by
\begin{equation}
	\label{eq:proc_dikta}
	\bar{R}_n^1(u) = \frac{1}{\sqrt{n}} \sumin \qty(Y_i - m(X_i, \betan)) \ind{\betan^T X_i \le u}, \quad u \in \bar{\R}.
\end{equation}
It is applicable to parametric and semi-parametric GLMs in which the regression function $m(\cdot, \cdot)$ is assumed to belong to a parametric family $\{x \mapsto m(x, \beta) = g^{-1}(\beta^T x) \,|\, \beta \in \R^p \}$ with $g$ denoting the invertible link function. Technically, this constitutes a test for the conditional mean -- the assumed conditional distribution function is solely used to calculate an estimator for $\betan$.

A different technique is used in the test procedure proposed in \cite{delgado2008distribution}. Here, the test statistic is derived from an empirical process of the Rosenblatt transformations $(U_i, V_i) = (\hat{H}_n(X_i), F(Y_i|\vtn, X_i))$ which follow a uniform distribution in the limit. To obtain an asymptotically distribution-free test statistic, the authors suggest performing a martingale transform. Although the paper is very extensive, also describing the possibility of performing directional tests, this approach is fairly involved, model-dependent and difficult to automate using software.

Yet another possible way to test for $H_0$, described in \cite{bierens2012integrated}, exploits the idea of comparing the empirical conditional characteristic function with the one implied by the model. The underlying process is defined as
\begin{equation}
	\label{eq:proc_bierens}
	Z_n(\tau, \xi) = \frac{1}{\sqrt{n}} \sum_{j=1}^n \qty(\exp\qty(i \tau Y_j) - \int \exp\qty(i \tau y) dF(y|\vtn, X_j)) \exp\qty(i \xi^T X_j).
\end{equation}
An adequate choice of the index set for this process, as well as a technique to avoid integration, is thoroughly discussed in the paper, ultimately resulting in the so-called simulated integrated conditional moment test. A drawback of this approach is the requirement to select an appropriate hyperparameter, denoted as $c$ in their paper, which defines the integration domain and can impact the small sample power of the test.

Finally, in \cite{veazie2020simple}, a modified Pearson Chi-square test using the uniformly distributed transformations $F(Y_i|\vtn, X_i)$ as inputs is suggested. This method also involves the choice of a hyperparameter, specifically the number of intervals $K$. Furthermore, its theoretical foundation is limited, as, for instance, consistency has not been proven.
Most of the stated methods involve the approximation of the critical value by a bootstrap method and many of them were shown to be consistent. Some of them were even extended to time series data or function-valued parameters (see \citet{bai2003testing, rothe2013misspecification, troster2021specification}).

In this paper, we derive a new test statistic for the problem at hand. Its advantages over the described methods proposed in the literature are manifold:
\begin{itemize}
	\item It is model-independent and thus easy to automate.
	\item It does not involve any hyperparameters, making it more robust for practical usage.
	\item It has a rigorous theoretical foundation.
	\item It is applicable for high-dimensional input vectors $X$.
\end{itemize}
Note that, as explained above, all other methods lack at least one of these qualities.

\Fref{sec:def} comprises a detailed derivation of the proposed test statistic. A result on the limit distribution of the underlying process is established in \fref{sec:asymp} which, in theory, can be used to approximate the $p$-value for large sample sizes $n$. However, it turns out that the limit distribution is dependent on the true distributions of $X$ and $Y$, which are unknown. To circumvent this problem and still be able to approximate the $p$-value, we suggest a parametric bootstrap method and establish its asymptotic justification in \fref{sec:boot}. The consistency of the resulting goodness-of-fit test is verified in \fref{sec:cons}. In \fref{sec:appl}, the finite sample behavior of our method is studied, applying the method to both simulated and real datasets and comparing the results to methods from the literature. Importantly, it can be seen that our new test seems to be more sensitive to deviations from the distribution family. Beyond that, we developed an R package called gofreg that streamlines the application of bootstrap-based goodness-of-fit tests for parametric regression and thereby enhances their usability. The appendix finally provides proofs of the theoretical results stated in the text.

\section{Definition of the test statistic}
\label{sec:def}
Our test statistic will be based on the difference between a non-parametric and a semi-parametric estimate of the marginal distribution function $F_Y$ of $Y$. A natural choice for the non-parametric estimator is the empirical distribution function (ecdf) $\hat{F}_{Y,n}$ of $Y_1, \dots, Y_n$. For the derivation of the estimator of $F_Y$ taking the assumed parametric family $\F$ for the conditional density function $f$ into account, we write
\begin{align*}
	F_Y(t) &= \expec{\ind{Y \le t}} = \expec{\expec{\ind{Y \le t} \,|\, X}}\\
	&= \int \expec{\ind{Y \le t} \,|\, X=x} H(dx)\\
	&= \int F(t|x) H(dx),
\end{align*}
where $F(\,\cdot\,|x)$ is the true conditional distribution function of $Y$ given \mbox{$X=x$}. The semi-parametric estimate of $F_Y$ now follows in two approximation steps. First, we substitute the true conditional distribution with a parametric estimate, yielding
\begin{align*}
	F_{Y, \vtn}(t) \coloneqq \int F\big(t|\vtn, x\big) H(dx).
\end{align*}
A classical choice for $\vtn$ is the maximum likelihood estimator (MLE). Our analysis is, however, worded in general terms and only requires the estimator to meet certain assumptions. As in most applications the distribution of the covariates is unknown, we further approximate $H$ by the ecdf $\hat{H}_n$ of $X_1, \dots, X_n$, resulting in
\begin{align*}
	\hat{F}_{Y,\vtn}(t) \coloneqq \int F(t|\vtn, x) \hat{H}_n(dx) = \frac{1}{n} \sumin F\big(t|\vtn, X_i\big).
\end{align*}
Now, we define the conditional empirical process with estimated parameters as
\begin{align*}
	\tnproc(t) &\coloneqq \sqrt{n} \left(\hat{F}_{Y,n}(t) - \hat{F}_{Y, \vtn}(t)\right) = \frac{1}{\sqrt{n}} \sumin \qty(\textcolor{black}{\ind{Y_i \le t} - F\big(t|\vtn, X_i\big)}), \quad t \in \bar{\R}.
\end{align*}

As a test statistic, we can use some continuous functional of the process $\tnproc$. The supremum norm $\norm{\tnproc} = \sup_{t \in \bar{\R}} \abs{\tnproc(t)}$, for example, yields a Kolmogorov-Smirnov type statistic, whereas the integral $\int \tnproc^2(t) F\big(t|\vtn, dx\big)$ represents a Cram\'er-von-Mises type statistic. In the following, we will consider $\norm{\tnproc}$ in particular, but the results can be easily transferred to other statistics. The $p$-value corresponding to an observed test statistic value $A$ is given by 
$$p \coloneqq \P_{H_0}(\norm{\tnproc} \ge A).$$
As usual, the null hypothesis is rejected if the $p$-value is below some level of significance. In order to be able to compute the $p$-value (or equivalently the critical value), we need to know the distribution of $\norm{\tnproc}$ under the null hypothesis. 

\section{Asymptotic null distribution}
\label{sec:asymp}

To investigate the asymptotic null distribution of our test statistic, we need to impose some conditions on the parametric model $\F$ and the estimator $\vtn$. We first state a set of assumptions that is closely aligned with the one in \citet{andrews1997conditional}. Later, we show that the result remains valid under an alternative set of conditions that is easier to verify in practice.

\begin{assumption}{M1}
	\label{ass:m1}
	Define $w(t, \vt, x) \coloneqq \frac{\del}{\del \vt} F(t|\vt,x)$ and $W(t, \vt) = \expec{w(t, \vt, X)}$.
	\begin{enumerate}[(i)]
		\item \label{cond:F_diff}
		There exists a neighborhood $V \subseteq \Theta$ of $\vt_0$ such that $\vt \mapsto F(t|\vt, x)$ is differentiable on $V$ for all $t$ and $H-$a.e.\ $x$.
		\item \label{cond:w_unif_wlln}
		For all non-negative sequences $\{r_n\}_{n \ge 1}$ with $r_n \to 0$ as $n \to \infty$, it holds
		$$\sup_{t \in \bar{\R}} \sup_{\norm{\vt - \vt_0} \le r_n} \norm{\frac{1}{n} \sumin w(t, \vt, X_i) - W(t, \vt_0)} \xrightarrow[n \to \infty]{} 0 \quad \text{in pr.}$$
		\item \label{cond:W_unif_cont}
		$W(t, \vt_0)$ is uniformly continuous in $t$. 
	\end{enumerate}
\end{assumption}
\pagebreak
\begin{assumption}{E1}
	\label{ass:e1}
	\hfill
	\begin{enumerate}[(i)]
		\item \label{cond:mle_lin}
		There exists a function $L$ such that
		$$\sqrt{n}(\vtn - \vt_0) = n^{-1/2} \sumin L(X_i,Y_i;\vt_0) + \opI. $$
		\item \label{cond:L_e_cov}
		$\expec{L(X,Y;\vt_0)} = 0$ and $\expec{L(X,Y;\vt_0)(L(X,Y;\vt_0))^T} < \infty$.
	\end{enumerate}
\end{assumption}

The estimator $\vtn$ admitting an asymptotic linear representation as assumed in \ref{ass:e1} is a classic condition for convergence theorems of parametric test statistics. Usually, it is fulfilled for least squares or maximum likelihood estimators (MLEs). For the MLE, a corresponding result was established in \citet[Corollary 5.56]{dikta2021bootstrap}. Technically, their setting was a little different in that they were considering parametric GLMs in particular, but the results can be easily extended to general conditional distribution families. 

The following theorem establishes a convergence result for the conditional empirical process with estimated parameters $\tnproc$. Specifically, we consider weak convergence in the space of uniformly bounded functions on the extended real line $\lspace$ as defined in \citet[Chap.\ 7]{kosorok2008introduction}. The asymptotic distribution of the test statistic $\norm{\tnproc}$ can be derived subsequently using the Continuous Mapping Theorem.

\begin{theorem}
	\label{thm:conv_proc}
	Under $H_0$ and Assumptions \ref{ass:m1} and \ref{ass:e1}, it holds that
	\begin{align*}
		\tnproc \Rightarrow \tinfproc \quad\text{in } \lspace,
	\end{align*}
	where $\tinfproc$ is a centered Gaussian process with covariance function
	\begin{align*}
		K(s,t) = &~ \prob[\vt_0]{Y \le \min(s,t)} - \expec{F(s|\vt_0,X) F(t|\vt_0,X)}\\
		&+ W^T(t,\vt_0) \expec[\vt_0]{\qty(F(s|\vt_0,X) - \ind{Y \le s}) L(X,Y;\vt_0)}\\
		&+ W^T(s,\vt_0) \expec[\vt_0]{\qty(F(t|\vt_0,X) - \ind{Y \le t}) L(X,Y;\vt_0)}\\
		&+ W^T(s,\vt_0) \cov[\vt_0]{L(X,Y;\vt_0)} W(t,\vt_0).
	\end{align*}
\end{theorem}

In fact, a similar result can be directly obtained from \cite[Theorem 1]{andrews1997conditional} using the Continuous Mapping Theorem. However, we want to emphasize that our Assumption \ref{ass:m1} is weaker than its analogue in Andrews' work. While the two sets of assumptions are largely comparable, the key distinction lies in the additional indicator function $\ind{X \le x}$ appearing in Andrews' formulation of $w(t, \vt, x)$. In particular, his assumptions analogous to \ref{ass:m1}\eqref{cond:w_unif_wlln} and \eqref{cond:W_unif_cont} have to hold uniformly for H-a.e.\ $x$, whereas we do not impose that requirement. It should further be noted that, as opposed to Andrews, we only need the convergence in \fref{ass:m1}\fref{cond:w_unif_wlln} to hold in pr.\ to prove \fref{thm:conv_proc}. The stronger assumption of convergence wp1 is needed to derive the asymptotic distribution of the bootstrap process in \fref{thm:conv_proc_boot}. The same applies to the uniform boundedness of $t \mapsto W(t, \vt_0)$.

Next, we introduce a different set of assumptions under which the conclusion of \fref{thm:conv_proc} remains valid. Although they are more restrictive, it is instructive to mention them as they are easier to verify and better comparable to the assumptions in \citet{dikta2021bootstrap}.

\begin{assumption}{M1'}
	\label{ass:m1p}
	Define $v(t, \vt, x) \coloneqq \frac{\del}{\del \vt} f(t|\vt,x)$.
	\begin{enumerate}[(i)]
		\item \label{cond:v_ex_dom}
		There exists a neighborhood $V \subseteq \Theta$ of $\vt_0$ in which $v(t, \vt, x)$ is defined and there is a function $g(t,x)$ such that $\norm{v(t,\vt,x)} \le g(t,x)$ and $\vt \in V$ with $\int g(t,x) \nu(dt) < \infty$ for $H$-a.e.\ $x$.
		\item \label{cond:v_ddom}
		For $V$ defined in (i), it holds
		$$ \int \sup_{\vt \in V} \int \norm{v(t,\vt,x)} \nu(dt) H(dx) < \infty. $$
		\item \label{cond:w_cont}
		The family of functions  $\{\vt \mapsto w(t,\vt,x)\}_{t \in \R}$ is equicontinuous at $\vt_0$ for $H$-a.e.\ $x$, meaning that for $H$-a.e.\ $x$ and every $\eps > 0$, there exists a $\delta \equiv \delta(x,\eps) > 0$ such that
		$$ \sup_{t \in \R} \norm{w(t, \vt, x) - w(t, \vt_0, x)} < \eps \quad\text{if}\quad \norm{\vt - \vt_0} < \delta. $$
		\item \label{cond:W_cont}
		$W(t,\vt)$ is uniformly continuous in $t$ at $\vt_0$.
	\end{enumerate}
\end{assumption}

\begin{corollary}
	\label{cor:conv_proc}
	\Fref{thm:conv_proc} also applies under Assumptions \ref{ass:m1p} and \ref{ass:e1}, since \ref{ass:m1p} implies \ref{ass:m1}.
\end{corollary}

Assumptions \ref{ass:m1p}\fref{cond:v_ex_dom} and \fref{cond:v_ddom} are analogous to \citet[6.4.3(iv) and (v)]{dikta2021bootstrap}. In their book, the parametric regression function $m(\cdot, \cdot)$ plays the role of the conditional density $f(\cdot\,|\,\cdot, \cdot)$ in our analysis. A sufficient condition for \fref{ass:m1p}\fref{cond:v_ddom} is $\int \int g(t,x) \nu(dt) H(dx)$ to be finite with $g$ defined in \fref{ass:m1p}\fref{cond:v_ex_dom}.
\Fref{ass:m1p}\fref{cond:W_cont} is the analogue of \citet[6.5.2(xi)]{dikta2021bootstrap}.

As the asymptotic null distribution in \fref{thm:conv_proc} involves $\vt_0$ as well as the distribution of $X$, it is case dependent and cannot be tabulated. For that reason, we suggest a bootstrap method to approximate the limit distribution.

\section{Parametric bootstrap method}
\label{sec:boot}

The goal of bootstrap methods, in general, is to estimate the distribution of a given test statistic under the null hypothesis by generating new samples, introducing some type of randomness while sticking as close as possible to the original sample. In our case, this means that we want to estimate the distribution of $\norm{\tnproc}$ by generating new samples $\{(\Xs, \Ys)\}$ whose conditional distribution is guaranteed to belong to the conditional distribution family $\F$ while at the same time keeping the distributions as similar as possible to the original sample. These considerations lead to the following resampling scheme:
\begin{enumerate}[~(1)]
	\item Keep the covariates the same: $\Xs = X_i$.
	\item Generate new response variables $Y_{i,n}^*$ according to the estimated conditional distribution function $F\big(\cdot | \vtn, \Xs\big)$.
	\item Based on this new bootstrap sample, find an estimate $\vtns$ for the distribution parameter.
	\item Determine the bootstrap conditional empirical process with estimated parameters
	\begin{align*}
		\tsnproc(t) \coloneqq \sqrt{n} \left(F^*_{Y,n}(t) - \hat{F}_{Y,\vtns} (t) \right),
	\end{align*}
	where $F^*_{Y,n}(t)$ denotes the ecdf of $\{\Ys\}_{i=1}^n$.
\end{enumerate}  \texttt{}

The $p$-value $\P_{H_0}\qty(\norm{\tnproc} \ge A)$ is then approximated by $\probs{\norm{\tsnproc} \ge A}$ with $\P_n^*$ indicating the probability measure corresponding to the bootstrap random variables based on $n$ original observations.	This approach is justified if the bootstrap process $\tsnproc$ converges to the same limit distribution as $\tnproc$. To prove a corresponding result, some additional assumptions are needed. Again, we first consider assumptions similar to those in \citet{andrews1997conditional} and in a second step provide stronger ones that are easier to verify in practice.

\begin{assumption}{M2}
	\label{ass:m2}
	Let $w$ and $W$ be the functions defined in \fref{ass:m1}.
	\begin{enumerate}[(i)]
		\item \label{cond:w_unif_slln}
		For all non-negative sequences $\{r_n\}_{n \ge 1}$ with $r_n \to 0$ as $n \to \infty$, it holds
		$$\sup_{t \in \bar{\R}} \sup_{\norm{\vt - \vt_0} \le r_n} \norm{\frac{1}{n} \sumin
		w(t, \vt, X_i) - W(t, \vt_0)} \xrightarrow[n \to \infty]{} 0 \quad \text{wp1}.$$
		\item \label{cond:W_unif_bdd}
		The function $t \mapsto W(t, \vt_0)$ is uniformly bounded, i.e. $ \sup_{t \in \bar{\R}} \norm{W(t, \vt_0)} < \infty. $
	\end{enumerate}
\end{assumption}

\begin{assumption}{ME2} 
	\label{ass:me2}
	Let $L$ be the function defined in \fref{ass:e1}. The convergence
	$$\frac{1}{n} \sumin \expecs{\ell_k(X_i, \Ys, \vtn)} \longrightarrow \expec[\vt_0]{\ell_k(X, Y, \vt_0)}$$
	holds wp1 for the following functions (for all values of $s, t \in \bar{\R}$):
	\begin{enumerate}[(i)]
		\item \label{cond:conv_I}
		$\ell_1(x,y,\vt) = \abs{\ind{y \le s} - \ind{y \le t}}$,
		\item \label{cond:conv_I-F}
		$\ell_2(x,y,\vt) = \qty(\ind{y \le s} - F(s|\vt, x)) \qty(\ind{y \le t} - F(t|\vt, x))$,
		\item \label{cond:conv_LL}
		$\ell_3(x,y,\vt) = L(x,y,\vt) \qty(L(x,y,\vt)^T)$, and
		\item \label{cond:conv_(I-F)L}
		$\ell_4(x,y,\vt) = \qty(\ind{y \le s} - F(s|\vt, x)) L(x,y,\vt)$.
	\end{enumerate}
\end{assumption}
\pagebreak

\begin{assumption}{E2}
	\label{ass:e2}
	Let $L$ be the function defined in \fref{ass:e1}.
	\begin{enumerate}[(i)]
		\item \label{cond:mle_lin_boot}
		$\sqrt{n}(\vtns - \vtn) = n^{-1/2} \sumin L(X_i,\Ys;\vtn) + \opIs.$ 
		\item \label{cond:xi_cov}
		$\expecs{L(X,Y^*;\vtn)} = 0.$
		\item \label{cond:L_delta}
		For the neighborhood $V$ defined in \fref{ass:m1}, there exists a $\delta > 0$ such that
		$$ \int \sup_{\vt \in V} \int \norm{L(x,y,\vt)}^{2+\delta} f(y|\vt,x) \nu(dy) H(dx) < \infty.$$
	\end{enumerate}
\end{assumption}

Assumptions \ref{ass:e2}\fref{cond:mle_lin_boot}, \ref{ass:e2}\fref{cond:xi_cov} and \ref{ass:me2}\fref{cond:conv_LL} are usually fulfilled for appropriate estimators $\vtn$. Their validity for the MLE can be proven similarly to \citet[Corollary~5.62]{dikta2021bootstrap}. 

The following theorem establishes a weak convergence result for the bootstrap process $\tsnproc$ and thereby justifies the bootstrap approximation.

\begin{theorem}
	\label{thm:conv_proc_boot}
	Under Assumptions \ref{ass:m1}, \ref{ass:m2}, \ref{ass:me2}, \ref{ass:e1}, \ref{ass:e2} and if $\vtn \to \vt_0$ wp1, it holds wp1 that
	\begin{align*}
		\tsnproc \Rightarrow \tinfproc \quad\text{in } \lspace,
	\end{align*}
	where $\tinfproc$ is the same process as in \fref{thm:conv_proc}.
\end{theorem}

Just as for \fref{thm:conv_proc}, a similar result could be directly obtained from \citet[Theorem 2]{andrews1997conditional}, but, importantly, this would entail stronger assumptions involving the indicator $\ind{X \le x}$. Note that \fref{ass:m2} is already included in \citet[Assumption M1]{andrews1997conditional}, and Assumptions \ref{ass:me2} and \ref{ass:e2} are the analogues of \citet[Assumptions M2(i) and E2]{andrews1997conditional}, respectively.

In a next step, we want to consider sufficient conditions that are easier to verify in practice.

\begin{assumption}{ME2'}
	\label{ass:me2p}
	Let $L$ be the function defined in \fref{ass:e1} and $V$ be the neighborhood defined in \fref{ass:m1p}.
	\begin{enumerate}[(i)]
		\item It holds that 
		$$ \int \int \sup_{\vt \in V} f(y|\vt,x) \nu(dy) H(dx) < \infty. $$
		\item \label{cond:xi_cont_int}
		$L$ is continuous in $\vt$ at $\vt_0$ and there exist neighborhoods $V_1$ and $V_2$ of $\vt_0$ such that
		$$ \int \int \sup_{\vt_1 \in V_1} \norm{L(x,y,\vt_1)} \sup_{\vt_2 \in V_2} f(y|\vt_2,x) \nu(dy) H(dx) < \infty. $$
	\end{enumerate}
\end{assumption}

\begin{corollary}
	\label{cor:conv_proc_boot}
	\Fref{thm:conv_proc_boot} also applies under Assumptions \ref{ass:m1p}, \ref{ass:me2p}, \ref{ass:me2}\fref{cond:conv_LL}, \ref{ass:e1} and \ref{ass:e2}.
\end{corollary}

In practice, the bootstrap $p$-value $\probs{\norm{\tsnproc} \ge A}$ is, in turn, approximated by a Monte Carlo simulation repeating steps (1)-(4) $m$ times yielding $\{\tilde{\alpha}_{n,k}^*\}_{k=1}^m$ and finally computing 
\begin{equation}
	\label{eq:p_approx}
	\tilde{p} \coloneqq \frac{1}{m} \sum_{k=1}^m \ind{\norm{\tilde{\alpha}_{n,k}^*} \ge A} \approx \probs{\norm{\tsnproc} \ge A} \approx \P_{H_0}\qty(\norm{\tnproc} \ge A) = p.
\end{equation}
This additional approximation is justified by the strong law of large numbers (SLLN).

Note that \fref{thm:conv_proc_boot} is also valid under $H_1$ as long as $\vtn$ converges to some $\vt_0 \in \Theta$ and Assumptions \ref{ass:m1}, \ref{ass:m2}, \ref{ass:me2}, \ref{ass:e1} and \ref{ass:e2} hold for this $\vt_0$. This result will be needed in the next section investigating the asymptotic power of the test.

\section{Consistency of the test}
\label{sec:cons}

In the following, we show that our proposed test is consistent against any conditional density $f_1$ in the alternative hypothesis $H_1$, as long as the corresponding conditional distribution function $F_1$ is distinguishable from all members of the parametric family.
\begin{assumption}{H1}
	\label{ass:h1}
	For every $\vt \in \Theta$, there exists a $t \in \R$ such that
	\begin{equation*}
		\expec{F_1(t|X)} \ne \expec{F(t|\vt,X)}.
	\end{equation*}
\end{assumption}
Further, we need to impose the following assumption on the parameter estimator $\vtn$.

\begin{assumption}{E3}
	\label{ass:e3}
	There exists a $\vt_1 \in \Theta$ such that $\vtn \to \vt_1$ in pr.
\end{assumption}

The limit value $\vt_1$ is sometimes called the pseudo-true parameter. In case $\vtn$ is the MLE, $\vt_1$ minimizes the Kullback-Leibler divergence between the true distribution of the data and the parametric model.

\begin{theorem}
	\label{thm:cons}
	Let $c_n$ denote the (random) sequence of bootstrap critical values for a given significance level $0 < \alpha < 1$, as defined by $\probs{\norm{\tsnproc} > c_n} \le \alpha$.
	Under the alternative hypothesis $H_1$, Assumptions \ref{ass:h1} and \ref{ass:e3} as well as  \ref{ass:m1}, \ref{ass:m2}, \ref{ass:me2}, \ref{ass:e1}, \ref{ass:e2} with $\vt_0$ being replaced by the pseudo-true value $\vt_1$, it holds that
	\begin{equation*}
		\lim_{n \to \infty} \prob{\norm{\tnproc} > c_n} = 1.
	\end{equation*}
\end{theorem}

\section{Simulations and examples}
\label{sec:appl}

In this section, we will investigate the finite sample behavior of the proposed goodness-of-fit test for parametric families of conditional distributions. For that, we will apply it to both artificially created data (for which the true distribution is known) and real datasets (for which the true distribution is unknown). The results will be compared to other established bootstrap-based tests used for the same purpose. Specifically, we will also evaluate the methods proposed by \citet{bierens2012integrated}, \citet{andrews1997conditional} and \citet{dikta2021bootstrap}, whose underlying processes are given in equations (\ref{eq:proc_andrews})-(\ref{eq:proc_bierens}). All tests were conducted in R using the gofreg-package, see \citet{gofreg}.

\vspace{0.5\baselineskip}
\noindent\textbf{Simulation studies} 
\vspace{0.5\baselineskip}

For our first simulation study, we use the same data-generating processes (DGPs) as in \citet{zheng2000consistent} that were also investigated in Bierens and Wang's paper. Accordingly, the one-dimensional covariate $X$ is sampled from a standard normal distribution, and the continuous response variable $Y$ is generated according to
\begin{description}
	\setlength\itemsep{0cm}
	\item[DGP(C0):] $Y = 1 + X + \eps$ where $\eps \sim N(0,1)$,
	\item[DGP(C1):] $Y = 1 + X + \eps$ where $\eps$ has the standard logistic distribution,
	\item[DGP(C2):] $Y = 1 + X + \eps$ where $\eps \sim t_5$,
	\item[DGP(C3):] $Y = 1 + X + X^2 + \eps$ where $\eps \sim N(0,1)$, or
	\item[DGP(C4):] $Y = 1 + X + X\eps$ where $\eps \sim N(0,1)$.
\end{description}
The null hypothesis $H_0$ is that the model is linear, homoscedastic and normally distributed. Under this assumption, DGP(C0) is true and all other DGPs are false. We want to analyze the sensitivity of the four goodness-of-fit tests to the different deviations from the null.

\begin{figure}[p]
	\centering
	\includegraphics[width=\textwidth]{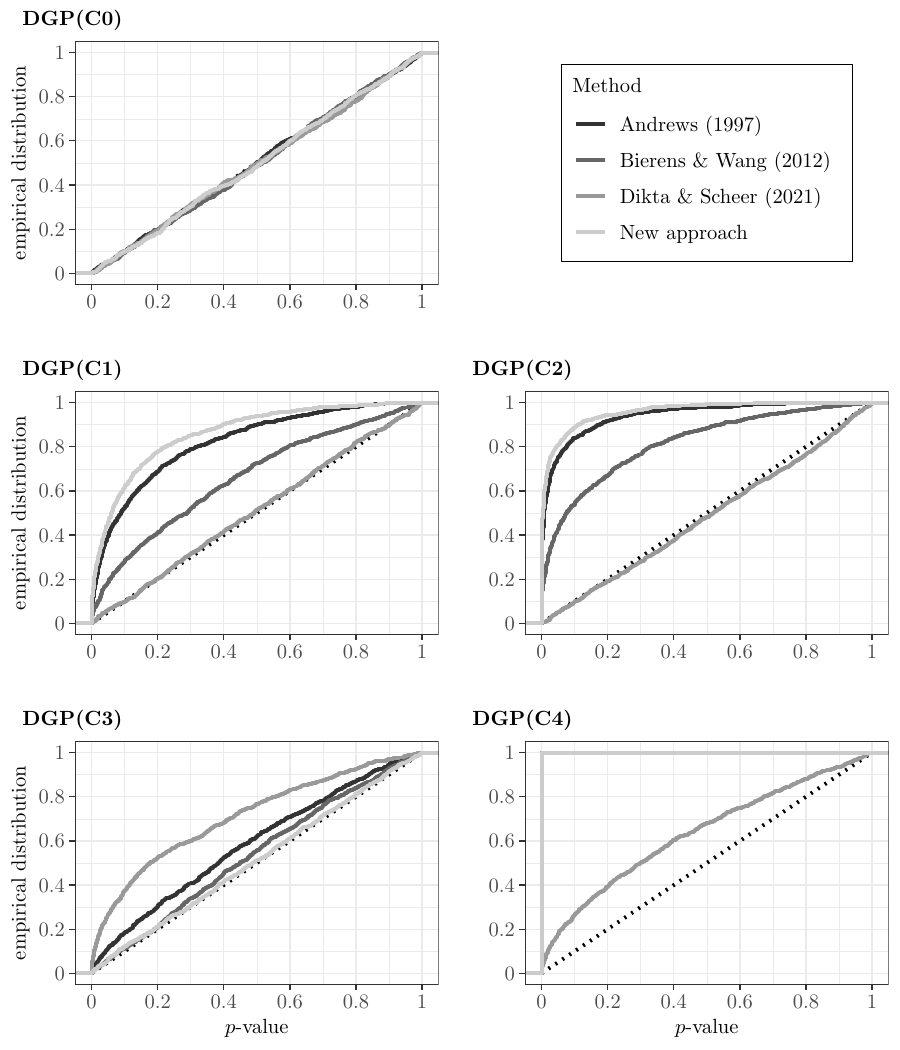}
	\caption{Empirical distribution of bootstrap $p$-values for different tests and simulated models with continuous response variable and sample size $n=500$.}
	\label{fig:gof_sim1}
\end{figure}

In all five simulations, we generated a dataset $\{(X_i,Y_i)\}_{i=1}^n$ with sample size $n=100$ or $n=500$ and applied the four bootstrap-based goodness-of-fit tests to it using $m = 500$ bootstrap repetitions. We always use the Kolmogorov-Smirnov type test statistic (based on the different underlying processes). Each method yields one $p$-value for the given dataset. To measure the sensitivity of each test, we repeated each simulation $1000$ times and computed the ecdf of the $p$-values -- in other words, the relative amount of rejections for different levels of significance. The corresponding results are illustrated in \Fref{fig:gof_sim1}. \Fref{tab:gof_sim1_n100} and \ref{tab:gof_sim1_n500} list the proportions of rejection corresponding to the common significance levels of $\alpha = 1\%$ and $\alpha = 5\%$.

Clearly, all four methods behave appropriately in case the null hypothesis is fulfilled. Specifically, the $p$-values for DGP(C0) are approximately uniformly distributed as the graphs are close to the dotted identity line, and the proportions of rejection roughly match the significance level. The plots corresponding to DGP(C1) and DGP(C2) indicate that our new approach is the most sensitive one to a violation of the distribution assumption. The likelihood of rejection (corresponding to low $p$-values) is the highest for our method, followed by Andrews' and Bierens and Wang's methods. The method studied by Dikta and Scheer, on the other hand, performs rather poorly in these examples as the $p$-value is approximately uniformly distributed. In the plot for DGP(C3), however, it can be seen that Dikta and Scheer's method seems to be the most sensitive one if the assumed linear relationship between $X$ and $Y$ is not valid. In this case, rejection is most likely using their method, whereas the other methods fail to reject $H_0$. Finally, all methods except Dikta and Scheer's react very sensitively to a violation of the homoscedasticity assumption, as is the case in DGP(C4). A comparison of \Fref{tab:gof_sim1_n100} and \ref{tab:gof_sim1_n500} shows that the results are robust to different sample sizes. If a test rejects $H_0$, the proportion of rejection is higher for $n=500$ than for $n=100$, which is expected.

\begin{table}[t]
	\centering
	\caption{Rejection percentage for different tests, significance levels and simulated models with continuous response variable and sample size $n=100$.}
	\label{tab:gof_sim1_n100}
	\vspace{0.25\baselineskip}
	\begin{tabular}[t]{lrrrrrrrrrr}
		\toprule
		\multicolumn{1}{c}{ } & \multicolumn{2}{c}{DGP(C0)} & \multicolumn{2}{c}{DGP(C1)} & \multicolumn{2}{c}{DGP(C2)} & \multicolumn{2}{c}{DGP(C3)} & \multicolumn{2}{c}{DGP(C4)} \\
		\cmidrule(l{3pt}r{3pt}){2-3} \cmidrule(l{3pt}r{3pt}){4-5} \cmidrule(l{3pt}r{3pt}){6-7} \cmidrule(l{3pt}r{3pt}){8-9} \cmidrule(l{3pt}r{3pt}){10-11}
		Method & 1\% & 5\% & 1\% & 5\% & 1\% & 5\% & 1\% & 5\% & 1\% & 5\%\\
		\midrule
		New approach & 1.1 & 5.6 & 4.9 & 14.3 & 13.4 & 26.3 & 1.2 & 5.9 & 84.6 & 96.1\\
		Andrews (1997) & 0.9 & 4.7 & 3.5 & 11.2 & 10.8 & 21.6 & 2.7 & 7.8 & 87.7 & 98.3\\
		Bierens \& Wang (2013) & 1.2 & 6.2 & 2.3 & 7.6 & 4.0 & 11.3 & 1.8 & 5.5 & 77.1 & 90.9\\
		Dikta \& Scheer (2021) & 0.8 & 4.8 & 0.7 & 4.0 & 0.7 & 4.6 & 2.3 & 9.2 & 5.3 & 15.5\\
		\bottomrule
	\end{tabular}
\end{table}

\begin{table}[t]
	\centering
	\caption{Rejection percentage for different tests, significance levels and simulated models with continuous response variable and sample size $n=500$.}
	\label{tab:gof_sim1_n500}
	\vspace{0.25\baselineskip}
	\begin{tabular}[t]{lrrrrrrrrrr}
		\toprule
		\multicolumn{1}{c}{ } & \multicolumn{2}{c}{DGP(C0)} & \multicolumn{2}{c}{DGP(C1)} & \multicolumn{2}{c}{DGP(C2)} & \multicolumn{2}{c}{DGP(C3)} & \multicolumn{2}{c}{DGP(C4)} \\
		\cmidrule(l{3pt}r{3pt}){2-3} \cmidrule(l{3pt}r{3pt}){4-5} \cmidrule(l{3pt}r{3pt}){6-7} \cmidrule(l{3pt}r{3pt}){8-9} \cmidrule(l{3pt}r{3pt}){10-11}
		Method & 1\% & 5\% & 1\% & 5\% & 1\% & 5\% & 1\% & 5\% & 1\% & 5\%\\
		\midrule
		New approach & 0.8 & 5.4 & 23.5 & 45.9 & 62.6 & 81.0 & 1.8 & 6.5 & 100 & 100\\
		Andrews (1997) & 1.6 & 5.1 & 17.7 & 39.7 & 54.4 & 75.4 & 3.3 & 11.6 & 100 & 100\\
		Bierens \& Wang (2013) & 0.6 & 5.0 & 7.3 & 18.6 & 22.9 & 42.9 & 1.8 & 6.4 & 100 & 100\\
		Dikta \& Scheer (2021) & 0.7 & 4.4 & 1.1 & 6.2 & 0.7 & 5.1 & 11.8 & 26.7 & 6.9 & 18\\
		\bottomrule
	\end{tabular}
\end{table}

\begin{figure}[p]
	\centering
	\includegraphics[width=\textwidth]{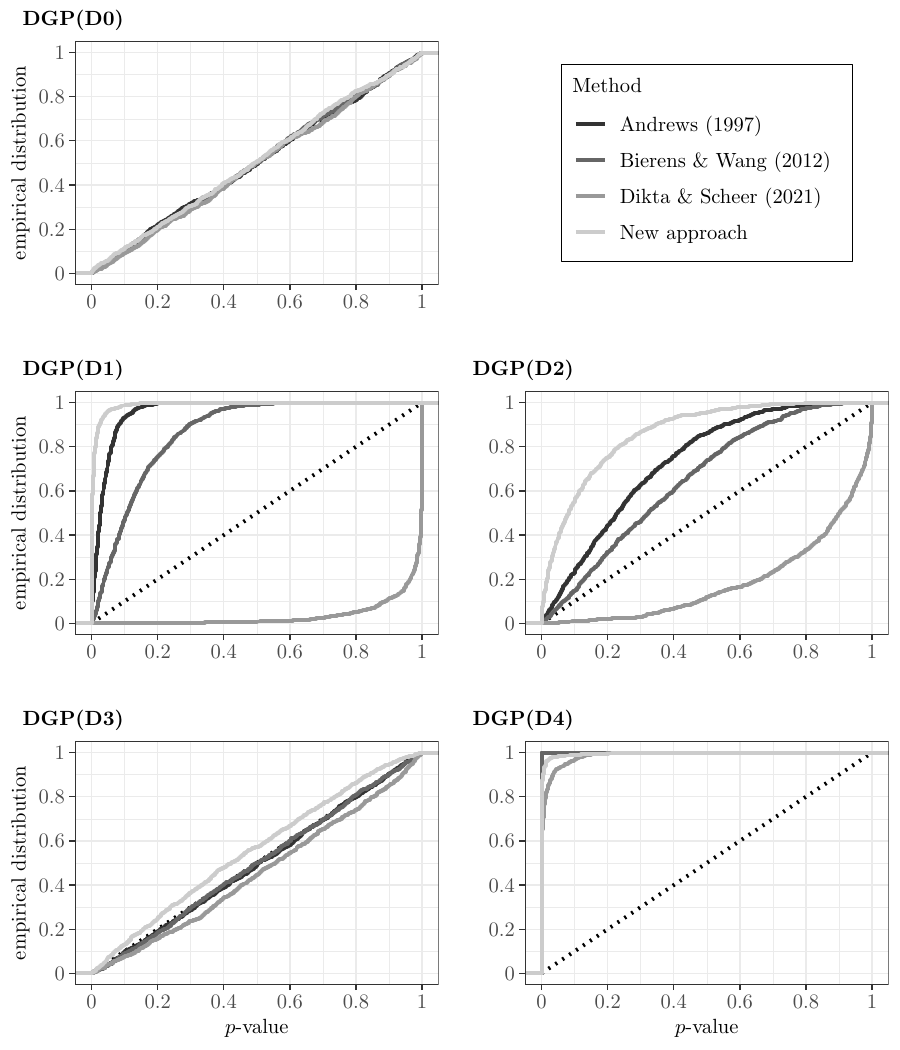}
	\caption{Empirical distribution of bootstrap $p$-values for different tests and simulated models with discrete response variable and sample size $n=500$.}
	\label{fig:gof_sim2}
\end{figure}

In the second simulation study, we consider a GLM with a discrete response variable. Specifically, we want to test for a Poisson distribution with a logarithmic link function, i.e. 
\begin{align*}
	H_0:Y|X \sim \text{Pois}\qty(\lambda = \exp(\beta^T X)).
\end{align*}
To create the samples, we use a standard normally distributed covariate $X$ and the following data-generating processes for the target variable $Y$:
\begin{description}
	\setlength\itemsep{0cm}
	\item[DGP(D0):] $Y|X \sim \text{Pois}\qty(\lambda = \exp(2 + 3X))$,
	\item[DGP(D1):] $Y|X \sim \text{Binom}\qty(n = \lceil 1.25 \cdot \exp(2 + 3X) \rceil,~ p = 0.8)$,
	\item[DGP(D2):] $Y|X \sim \text{Binom}\qty(n = \lceil 2 \cdot \exp(2 + 3X) \rceil,~ p = 0.5)$, 
	\item[DGP(D3):] $Y|X \sim \text{Binom}\qty(n = \lceil 10 \cdot \exp(2 + 3X) \rceil,~ p = 0.1)$, 
	\item[DGP(D4):] $Y|X \sim \text{NB}\qty(r = 0.25 \cdot \exp(2 + 3X),~ p = 0.2)$.
\end{description}
Note that $H_0$ only holds for DGP(D0). The parameters of the binomial and negative binomial distributions were chosen in a way that the conditional mean is the same in all five setups. We used the same procedure, sample size and repetition numbers as in the first simulation study. The results are illustrated in \Fref{fig:gof_sim2} as well as \Fref{tab:gof_sim2_n100} and \ref{tab:gof_sim2_n500}. 

\begin{table}[t]
\centering
\caption{Rejection percentage for different tests, significance levels and simulated models with discrete response variable and sample size $n=100$.}
\label{tab:gof_sim2_n100}
\vspace{0.25\baselineskip}
\small		
\begin{tabular}[t]{lrrrrrrrrrr}
	\toprule
	\multicolumn{1}{c}{ } & \multicolumn{2}{c}{DGP(D0)} & \multicolumn{2}{c}{DGP(D1)} & \multicolumn{2}{c}{DGP(D2)} & \multicolumn{2}{c}{DGP(D3)} & \multicolumn{2}{c}{DGP(D4)} \\
	\cmidrule(l{3pt}r{3pt}){2-3} \cmidrule(l{3pt}r{3pt}){4-5} \cmidrule(l{3pt}r{3pt}){6-7} \cmidrule(l{3pt}r{3pt}){8-9} \cmidrule(l{3pt}r{3pt}){10-11}
	Method & 1\% & 5\% & 1\% & 5\% & 1\% & 5\% & 1\% & 5\% & 1\% & 5\%\\
	\midrule
	New approach & 1.1 & 6.5 & 4.8 & 18.1 & 2.7 & 8.5 & 1.3 & 4.9 & 35.1 & 56.3\\
	Andrews (1997) & 1.6 & 6.5 & 0.5 & 3.2 & 0.4 & 2.7 & 1.1 & 5.1 & 69.6 & 87.8\\
	Bierens \& Wang (2013) & 1.0 & 4.0 & 0.4 & 2.6 & 0.5 & 3.2 & 1.6 & 5.8 & 86.3 & 95.8\\
	Dikta \& Scheer (2021) & 0.7 & 4.5 & 0.1 & 0.1 & 0.0 & 0.2 & 0.8 & 3.1 & 73.4 & 89.6\\
	\bottomrule
\end{tabular}
\end{table}

\begin{table}[t]
\centering
\caption{Rejection percentage for different tests, significance levels and simulated models with discrete response variable and sample size $n=500$.}
\label{tab:gof_sim2_n500}
\vspace{0.25\baselineskip}
\small	
\begin{tabular}[t]{lrrrrrrrrrr}
	\toprule
	\multicolumn{1}{c}{ } & \multicolumn{2}{c}{DGP(D0)} & \multicolumn{2}{c}{DGP(D1)} & \multicolumn{2}{c}{DGP(D2)} & \multicolumn{2}{c}{DGP(D3)} & \multicolumn{2}{c}{DGP(D4)} \\
	\cmidrule(l{3pt}r{3pt}){2-3} \cmidrule(l{3pt}r{3pt}){4-5} \cmidrule(l{3pt}r{3pt}){6-7} \cmidrule(l{3pt}r{3pt}){8-9} \cmidrule(l{3pt}r{3pt}){10-11}
	Method & 1\% & 5\% & 1\% & 5\% & 1\% & 5\% & 1\% & 5\% & 1\% & 5\%\\
	\midrule
	New approach & 2.4 & 5.9 & 80.0 & 96.1 & 15.2 & 38.8 & 1.2 & 7.2 & 94.0 & 98.3\\
	Andrews (1997) & 1.5 & 5.3 & 23.9 & 73.4 & 2.8 & 12.0 & 1.1 & 4.1 & 100.0 & 100.0\\
	Bierens \& Wang (2013) & 1.0 & 5.3 & 4.2 & 25.7 & 1.1 & 7.5 & 0.6 & 3.6 & 100.0 & 100.0\\
	Dikta \& Scheer (2021) & 1.2 & 4.3 & 0.0 & 0.0 & 0.0 & 0.3 & 0.6 & 4.2 & 79.3 & 93.0\\
	\bottomrule
\end{tabular}
\end{table}

All testing methods behave properly in case of  DGP(D0), where the model assumption is correct. Comparing the results for the different binomial distributions, namely DGP(D1)-DGP(D3), it can be seen that it becomes increasingly difficult to detect a deviation from the null hypothesis. Taking into account that the binomial distribution approaches a Poisson distribution as the number of trials $n$ tends to infinity while the mean stays constant, this behavior can be expected.
Importantly, it can be seen that our new test demonstrates the strongest rejection of the incorrect model for DGP(D1) and DGP(D2), significantly outperforming the other methods. The tests proposed by \cite{bierens2012integrated} and \cite{dikta2021bootstrap} fail to detect the deviation from the model assumption in this specific setup. It should be noted, though, that Dikta and Scheer's test is meant to test for the regression function (not the entire conditional distribution), which is correctly specified here. Interestingly, the comparison of the different testing methods changes for DGP(D4) with an underlying negative binomial distribution. In this case, all four methods clearly reject the model assumption, but this time, the tests from \cite{bierens2012integrated} and \cite{andrews1997conditional} show the highest proportion of rejection.

To investigate the power of the different methods in more detail and thereby shed light on the cases in which each is most sensitive, a theoretical comparison of the variances of the respective limit processes would be informative. However, as such an analysis is a more involved endeavor, we leave it for future research. Here, we merely want to give some first insights into the comparison between our method and that of Andrews. The variance of the process $\tinfproc$ is given in \fref{thm:conv_proc} as
\begin{align}
	\text{Var}\qty(\tinfproc(t)) &= K(t,t) \nonumber\\
	&= \prob[\vt_0]{Y \le t} - \expec{\qty(F(t|\vt_0, X))^2} \nonumber\\
	&\quad+ 2 W^T(t,\vt_0) \expec{\qty(F(t|\vt_0,X) - \ind{Y \le t}) L(X,Y;\vt_0)} \nonumber\\
	&\quad+ W^T(t, \vt_0) \cov{L(X,Y;\vt_0)} W(t,\vt_0). \label{eq:var_ours}
\end{align}
In contrast, the process $\nu_n(t,x)$ defined in \fref{eq:proc_andrews} and analyzed in \cite{andrews1997conditional} converges to a mean zero Gaussian process $\nu_\infty(t,x)$ with variance 
\begin{align}
	\text{Var}\qty(\nu_\infty(t,x)) &= \prob[\vt_0]{X \le x, Y \le t} - \expec{\qty(F(t|\vt_0, X))^2 \ind{X \le x}} \nonumber\\
	&\quad+2W_\nu^T(t, x, \vt_0) \expec{\qty(F(t|\vt_0,X) - \ind{Y \le t}) L(X,Y;\vt_0) \ind{X \le x}} \nonumber\\
	&\quad+W_\nu^T(t, x, \vt_0) \cov{L(X,Y;\vt_0)} W_\nu(t, x, \vt_0) \label{eq:var_andrews}
\end{align}
where $W_\nu(t, x, \vt_0) = \expec{w(t,\vt_0,X) \ind{X \le x}}$. A general comparison of these variances is not particularly meaningful, as for example 
$$\prob[\vt_0]{X \le x, Y \le t} \le \prob[\vt_0]{Y \le t}$$ 
while
$$- \expec{\qty(F(t|\vt_0, X))^2 \ind{X \le x}} \ge - \expec{\qty(F(t|\vt_0, X))^2}.$$
Moreover, to compare the third and fourth summands of the variances in \eqref{eq:var_ours} and \eqref{eq:var_andrews}, the relationship between $W(t, \vt_0)$ and $W_\nu(t,x,\vt_0)$ has to be investigated. Since 
$$W(t,\vt_0) = \expec{w(t,\vt_0, X) \qty(\ind{X \le x} + \ind{X > x})} = W_\nu(t,x,\vt_0) + \expec{w(t,X,\vt_0) \ind{X>x}},$$
the dominance of one term or the other depends on the sign of $w(t,x,\vt_0) = \frac{\del}{\del \vt} F(t|\vt_0,x)$, which can vary with $x$. Thus, a more informative comparison would likely require focusing on specific values of $x$.

\vspace{0.5\baselineskip}
\noindent\textbf{Bank transaction data} 
\vspace{0.5\baselineskip}

As a real-world example, we use the \texttt{Transact} dataset from the \textit{car} package in R, which contains the transaction times from $n=261$ branch offices of a large Australian bank \citep{cunningham1989estimating, fox2018r}. 
The data are composed of three variables: t1 and t2, counting the number of transactions of two different types, and time, the total time of labor needed to process the transactions. Some summary statistics of the data are given in \Fref{tab:bank_data}. In our analysis, we use $X = (\text{t1}, \text{t2})$ and $Y = \text{time}$.

\begin{table}[t]
	\caption{Description of variables in the bank transaction dataset.}
	\label{tab:bank_data}
	\vspace{0.5\baselineskip}
	\centering
	\begin{tabular}{lp{7cm}rrrr}
			\toprule
			Variable & Description & Min & Max & Mean & Median \\
			\midrule
			t1 & Number of type 1 transactions & 0& 1450 & 281.2 & 214\\
			t2 & Number of type 2 transactions & 148 & 5791 & 2422 & 2192\\
			time & Total transaction time in minutes & 487 & 20741 & 6607 & 5583\\
			\bottomrule
	\end{tabular}
\end{table}

In a first step, we consider a classical linear model with normal distribution, i.e. 
$$Y = \beta_0 + \beta_1 X_1 + \beta_2 X_2 + \eps, \quad \eps \sim \mathcal{N}(0,\sigma^2),$$
or, equivalently, in our notation,
$$\vt \equiv (\beta, \sigma) \in \R^3 \times \R, \quad F(\cdot|\vt, x) \equiv \mathcal{N}(\beta_0 + \beta_1 x_1 + \beta_2 x_2, \sigma^2).$$ 
We estimate $\vt$ via MLE and compute the $p$-values for $H_0$ according to the three different goodness-of-fit tests using $m=500$ bootstrap replications. The results are shown in \Fref{tab:bank_pvals}. The method introduced by \citet{dikta2021bootstrap} yields a $p$-value of $0.946$, so the model is clearly accepted. The goodness-of-fit tests from \citet{andrews1997conditional} and \citet{bierens2012integrated} as well as our new approach, however, reject the model with a $p$-value of $0.002$, $0.014$ and $0.088$, respectively. In light of our simulation studies, we could conclude that the regression function is probably correct (because Dikta and Scheer's method accepts the models), but the distribution family is not chosen appropriately (because the other three methods reject the models).

\begin{table}[t]
	\label{tab:bank_pvals}
	\centering
	\caption{Results for the bank transaction dataset. Left: MLE of the distribution parameter $\vt$ for the two fitted models. Right: Bootstrap $p$-values of different tests for the two fitted models.}
	\vspace{0.5\baselineskip}
	
	\begin{minipage}{0.38\textwidth}
		\centering
		\begin{tabular}{crr}
			\toprule
			Variable & Gaussian & Gamma\\
			\midrule
			$\beta_0$ & 144.70 & 125.85\\
			$\beta_1$ & 5.46 & 5.71\\
			$\beta_1$ & 2.03 & 2.01\\
			$\sigma$ & 1136.14 & 35.08\\
			\bottomrule
		\end{tabular}
	\end{minipage}
	\begin{minipage}{0.6\textwidth}
		\centering
		\begin{tabular}{lcc}
			\toprule
			Method &Gaussian &Gamma\\
			\midrule
			New approach & 0.088 &0.868\\
			Andrews (1997) &0.002 &0.418\\
			Bierens \& Wang (2012) &0.014 &0.346\\
			Dikta \& Scheer (2021) &0.946 &0.992\\
			\bottomrule
		\end{tabular}
	\end{minipage}	
\end{table} 

To come up with a more appropriate distribution family for $Y$ given $X$, we plotted the data and examined how the points are scattered around the mean. It could be seen that the variance of the data points does not seem to be constant, as it would be the case in a Gaussian model. Instead, the data points are closer around the regression line for lower values of the conditional mean and more spread out for higher values. This behavior suggests a Gamma distribution, implying a constant coefficient of variation.

So, in a next step, we consider a linear model with a Gamma distribution, i.e.
$$ \vt \equiv (\beta, \sigma) \in \R^3 \times \R, \quad F(\cdot|\vt, x) \equiv \text{Gamma}(\text{scale} = (\beta_0 + \beta_1 x_1 + \beta_2 x_2)/\sigma,\, \text{shape} = \sigma), $$
such that, as before,
$$\expec{Y|X=x} = \beta_0 + \beta_1 x_1 + \beta_2 x_2.$$
As before, we fit the model to the dataset via MLE and use the three different tests to evaluate its goodness-of-fit. As illustrated in \Fref{tab:bank_pvals}, all of the methods yield high $p$-values this time, meaning that the model is not rejected and thus seems to describe the data appropriately.

\vspace{0.5\baselineskip}
\noindent\textbf{Bike sharing data} 
\vspace{0.5\baselineskip}

As another real-world dataset, we consider the bike sharing data that were also analyzed in \citet{dikta2021bootstrap}. It was first considered in \citet{fanaee2014event} and can be downloaded from the UC Irvine Machine Learning Repository \citep{bikesharing}. The dataset contains the daily count of rental bikes in Washington DC in 2011 and 2012, together with corresponding weather and seasonal information. The data are preprocessed in the same way as discussed in \citet[Example 5.44]{dikta2021bootstrap}, leaving us with continuous variables for the normalized temperature, humidity and wind speed as well as factors indicating the weather situation, year, season and type of day. A more detailed description of the variables, including some summary statistics, is given in \Fref{tab:bike_data}.

\begin{table}[t]
	\caption{Description of variables in the bike sharing dataset after preprocessing.  Table split into continuous and discrete variables.}
	\label{tab:bike_data}
	\vspace{0.5\baselineskip}
	\centering
	\begin{tabular}{l>{\raggedright}p{8cm}rrrr}
		\toprule
		Variable & Description & Min & Max & Mean & Median \\
		\midrule
		registered & Count of registered users on a given day & 416 & 6946 & 3635 & 3603\\
		temp & Normalized temperature in Celsius & 0.6 & 0.86 & 0.51 & 0.53\\
		hum\_imp & Normalized humidity with missing values replaced by average of that month & 0.19 & 0.97 & 0.63 & 0.63\\
		windspeed & Normalized wind speed & 0.02 & 0.51 & 0.19 & 0.18\\
		\bottomrule
	\end{tabular}
	
	\vspace*{0.5\baselineskip}
	\begin{tabular}{l>{\raggedright}p{8cm}rrrrr}
		\toprule
		Variable & Description & \#0 & \#1 & \#2 & \#3 & \#4 \\
		\midrule
		year & Year (0: 2011, 1: 2012) & 365 & 302 &-&-&-\\
		season & Season (1: Spring, 2: Summer, 3: Fall,\\\qquad\quad~~4: Winter) &-& 170 & 184 & 188 & 125\\
		workingday & 1: Day is neither weekend nor holiday,\\ 0: Otherwise & 210 & 457 &-&-&-\\
		weathersit & 1: Clear/few clouds/partly cloudy,\\2: Mist + cloudy/broken clouds/few clouds, 3: Light snow/light rain + scattered clouds\\\qquad\qquad\qquad (+ thunderstorm),\\ 4: Heavy rain + ice pallets + thunderstorm\\\quad\, + mist/snow + fog  & - & 429 & 219 & 19 & 0\\
		holiday & 1: Day is holiday, 0: Otherwise & 649 & 18 &-&-&-\\
		christmas & 1: Day is between Christmas and New Year, 0: Otherwise &  659 & 8 &-&-&-\\
		\bottomrule
	\end{tabular}
\end{table}

As Dikta and Scheer, we use the daily rental counts (registered) as the output variable $Y$ and all other listed variables together with an intercept, the squared temperature, the squared humidity and an interaction term between year and season as the covariate vector $X$. In section 6.1.2 of their book, Dikta and Scheer identify two parametric GLMs---a negative binomial (NB) and a log-transformed Gaussian (LTG) model---that appear to fit the given data, as they are not rejected by their goodness-of-fit test at a significance level of $0.05$.
In each model, the canonical link function is used: the logarithm for the negative binomial model and the identity function for the Gaussian model. In particular, in NB, it is assumed that $Y = \text{registered}$ and 
$$\vt \equiv (\beta, r) \in \R^{17} \times \R, \quad F(\cdot|\vt, x) \equiv \text{NB}\qty(r = r, p = \frac{r}{r + \exp(\beta^T x)})$$
such that
$$ \expec{Y|X=x} = \exp(\beta^T x). $$
The Gaussian model, on the other hand, is called ``log-transformed'' as it uses $Y = \log(\text{registered})$ as the response variable. In LTG, we consider
$$\vt \equiv (\beta, \sigma) \in \R^{17} \times \R, \quad F(\cdot|\vt, x) \equiv \mathcal{N}\qty(\beta^T x, \sigma^2).$$
The respective MLE values for $\vt$ are listed in \Fref{tab:bike_pvals}.

We want to investigate how well these two models fit the given data according to our new approach using the conditional empirical process with estimated parameters. Again, we use $m=500$ bootstrap replications to evaluate each of the four discussed tests. As illustrated in \Fref{tab:bike_pvals}, our new goodness-of-fit test results in an approximate $p$-value of zero for both parametric families, so they are clearly rejected. The methods from \citet{andrews1997conditional} and \citet{bierens2012integrated} also yield very low $p$-values. 
Recalling the results of the simulation studies, such a combination is likely caused by a correct regression function but an inappropriate distribution assumption in the model.

\begin{table}[p]
	\centering
	\caption{Results for the bike sharing dataset. Left: MLE of the distribution parameter $\vt$ for the two fitted models. Right: Bootstrap $p$-values of different tests for the two fitted models.}
	\label{tab:bike_pvals}
	\vspace{0.5\baselineskip}
	\begin{minipage}{0.43\textwidth}
		\centering
		\begin{tabular}{crr}
				\toprule
				Coefficient of & LTG & NB\\
				\midrule
				(Intercept) & 6.15 & 6.17\\
				temp & 4.23 & 4.19\\
				$\text{temp}^2$ & -3.30 & -3.28\\
				hum\_imp & 1.16 & 1.16\\
				$\text{hum\_imp}^2$ & -1.35 & -1.32\\
				windspeed & -0.73 & -0.71\\
				yr1 & 0.71 & 0.70\\
				season2 & 0.37 & 0.37\\
				season3 & 0.45 & 0.44\\
				season4 & 0.54 & 0.53\\
				yr1 $\cdot$ season2 & -0.28 & -0.27\\
				yr1 $\cdot$ season3 & -0.29 & -0.28\\
				yr1 $\cdot$ season4 & -0.25 & -0.23\\
				workingday1 & 0.28 & 0.27\\
				weathersit2 & -0.07 & -0.07\\
				weathersit3 & -0.55 & -0.49\\
				holiday1 & -0.08 & -0.06\\
				christmas1 & -0.16 & -0.09\\
				($\sigma$ or $r$) & 0.18 & 34.90\\
				\bottomrule
		\end{tabular}
	\end{minipage}
	\begin{minipage}{0.55\textwidth}
		\centering
		\begin{tabular}{lcc}
			\toprule
			Method & LTG & NB\\
			\midrule
			New approach &0.000 &0.000\\
			Andrews (1997) &0.018 &0.024\\
			Bierens \& Wang (2012) &0.048 &0.000\\
			Dikta \& Scheer (2021) &0.142 &0.080\\
			\bottomrule
		\end{tabular}
	\end{minipage}

\end{table} 

Note that the high-dimensionality of the covariate vector $X$ results in a fairly long runtime for Andrews' method. In particular, the calculation of the $p$-value took about 8 times longer than for our new approach and roughly 20 times longer than using Dikta and Scheer's method.
\pagebreak

\appendix
\section{Proofs} %

\begin{proof}[Proof of \fref{thm:conv_proc}]
	The proof will be based on a Durbin-like splitting of the process (see \citet{durbin1973weak}) given by
	\begin{align*}
		\tnproc(t) &= \underbrace{\frac{1}{\sqrt{n}} \sumin \qty(\textcolor{black}{\ind{Y_i \le t} - F(t|\vt_0,X_i)})}_{\nproc{1}(t)}\\
		&\quad+ \underbrace{\frac{1}{\sqrt{n}} \sumin \qty(\textcolor{black}{F(t|\vt_0,X_i) - F(t|\vtn,X_i)})}_{\nproc{2}(t)}.
	\end{align*}
	To prove the convergence of $\tnproc$, we will first verify that the two processes are asymptotically tight in $\lspace$ and find asymptotic iid representations for them to analyze the covariance structure of the finite-dimensional distributions (fidis). Then we use the multivariate central limit theorem (CLT) for the convergence of the fidis and finally apply \citet[Theorem 7.17]{kosorok2008introduction} to conclude the weak convergence of the process in $\lspace$.
	
	Asymptotic tightness of $\nproc{1}$ will be shown by splitting it into two parts again:
	\begin{align*}
		\nproc{1}(t) &= \underbrace{\sqrt{n} \Big(\Fn(t) - \Fo(t)\Big)}_{\nproc{1a}(t)} + \underbrace{\sqrt{n} \left(\Fo(t) - \frac{1}{n} \sumin F(t|\vt_0,X_i)\right)}_{\nproc{1b}(t)}.
	\end{align*}
	The first summand $\nproc{1a}$ represents the classical empirical process. It was proven to be a Donsker class by Donsker himself, see e.g. \citet[p.\ 11]{kosorok2008introduction}. By \citet[Lemma 7.12(ii)]{kosorok2008introduction}, it follows that $\nproc{1a}$ is asymptotically tight in $\lspace$. 
	
	To show asymptotic tightness of the second summand $\nproc{1b}$, we write the process as 
	\begin{align*}
		-\nproc{1b}(t) &= \sqrt{n} \left(\frac{1}{n} \sumin F(t|\vt_0,X_i) - \Fo(t)\right)\\
		&= \sqrt{n} \left(\frac{1}{n} \sumin F(t|\vt_0,X_i) -\expec{F(t|\vt_0,X)}\right).
	\end{align*}
	This shows that $\nproc{1b}$ can be regarded as a generalized empirical process over the index set $\tilde{\F} = \{x \mapsto F(t|\vt_0,x) ~|~ t \in \R\}$. Now, we will use \citet[Theorem 8.19]{kosorok2008introduction} to show that $\tilde{\F}$ is a $\P$-Donsker class and consequently $\nproc{1b}$ is asymptotically tight. An envelope function of $\tilde{\F}$ is given by $\tilde{F}(x) = 1$, which is clearly square-integrable. 
	As explained in the first paragraph on page 150 of Kosorok's book, all measurability conditions of the theorem are satisfied if $\tilde{\F}$ is pointwise measurable. To see that this is in fact the case, define the countable set of functions $\tilde{\mathcal{G}} \coloneqq \{x \mapsto F(t|\vt_0,x) ~|~ t \in \Q\} \subset \tilde{\F}$. Due to the right-continuity of $F(t|\vt_0, x)$ in $t$, there exists a sequence $\{s_m\} \in \Q$ for every $t \in \R$ with $F(s_m|\vt_0,x) \to F(t|\vt_0,x)$ for every $x$. The final and main condition of \citet[Theorem 8.19]{kosorok2008introduction} is the boundedness of the uniform entropy integral. As noted at the beginning of page 158 in Kosorok's book, it is satisfied if $\tilde{\F}$ is a VC-class of functions. This is true since $F(t|\vt_0,x)$ is monotonically increasing in $t$ for every $x$, see \citet[Lemma 9.10]{kosorok2008introduction}.
	
	A linear representation of $\nproc{1}$ is given by
	\begin{align*}
		\nproc{1}(t) = n^{-1/2} \sumin \xi^{(1)}(t, X_i, Y_i)
	\end{align*}
	with 
	\begin{align*}
		\xi^{(1)}(t,x,y) &= \ind{y \le t} - F(t|\vt_0,x)\\
		\expec[\vt_0]{\xi^{(1)}(t,X,Y)} &= 0,\\
		\cov[\vt_0]{\xi^{(1)}(s,X,Y), \xi^{(1)}(t,X,Y)} %
		&= \prob[\vt_0]{Y \le \min(s,t)} - \expec{F(s|\vt_0,X) F(t|\vt_0,X)}
	\end{align*}
	
	Next, we want to examine $\nproc{2}(t)$.  Using \fref{ass:e1}\fref{cond:mle_lin} and the mean value theorem, we get
	\begin{align*}
		\nproc{2}(t) &= \frac{1}{\sqrt{n}} \sumin \qty(\textcolor{black}{F\big(t|\vt_0, X_i\big) - F\big(t|\vtn, X_i\big)})\\
		&= -\sqrt{n} \left(\vtn - \vt_0\right)^T \frac{1}{n} \sumin \frac{\del}{\del \vt} F(t|\tilde{\vt}_n(t,X_i), X_i)\\
		&= -\left(n^{-1/2} \sumin L^T(X_i,Y_i; \vt_0) + \opI\right) \left(\frac{1}{n} \sumin w(t,\tilde{\vt}_n(t,X_i),X_i)\right)
	\end{align*}
	with $\tilde{\vt}_n(t,X_i)$ lying on the line segment between $\vt_0$ and $\vtn$ dependent on $t$ and $X_i$. By \fref{ass:e1}\fref{cond:mle_lin} and the weak law of large numbers (WLLN), $\vtn$ converges to $\vt_0$ in pr. Thus, there exists a non-negative sequence $\{r_n\}_{n \ge 1}$ such that $r_n \to 0$ and $\prob{\norm{\vtn - \vt_0} > r_n} \to 0$ as $n \to \infty$.
	An application of the triangle inequality yields
	\begin{align*}
		&\prob{\sup_{t \in \bar{\R}} \norm{\frac{1}{n} \sumin w(t, \tilde{\vt}(t, X_i), X_i) - W(t, \vt_0)} > \eps}\\
		&\quad \le \prob{\sup_{t \in \bar{\R}} \sup_{\norm{\vt - \vt_0} \le r_n} \norm{\frac{1}{n} \sumin w(t, \tilde{\vt}(t, X_i), X_i) - W(t, \vt_0)} > \eps} + \prob{\norm{\vtn - \vt_0} > r_n},
	\end{align*}
	which converges to zero by the arguments above and \fref{ass:m1}\fref{cond:w_unif_wlln}.
	
	So we get
	\begin{align*}
		\nproc{2}(t) = n^{-1/2} \sumin \xi^{(2)}(t,X_i,Y_i) + r^{(2)}_n(t)
	\end{align*}
	with
	\begin{align*}
		\xi^{(2)}(t,x,y) &= -\left(L(x,y; \vt_0)\right)^T W(t, \vt_0)\\
		\expec[\vt_0]{\xi^{(2)}(t,X,Y)} &= 0,\\
		\cov[\vt_0]{\xi^{(2)}(s,X,Y), \xi^{(2)}(t,X,Y)} %
		&= W^T(s,\vt_0) \cov[\vt_0]{L(X,Y; \vt_0)} W(t,\vt_0)\\
		r^{(2)}_n(t) &\to 0 \text{ in pr.\ uniformly in } t.
	\end{align*}
	To prove the tightness of $\nproc{2}(t)$, note that $L$ is not dependent on $t$ and $W(t,\vt_0)$ is continuous in $t$ by assumption \fref{ass:m1}\fref{cond:W_unif_cont}. Thus, the process is in $\cspace$ and we can use \citet[Theorem 8.2]{billingsley1968convergence} to verify tightness. By the multivariate CLT and assumption \fref{ass:e1}\fref{cond:L_e_cov}, there is a centered normal random vector $S_\infty$ such that $S_n = -n^{-1/2} \sumin L(X_i,Y_i; \vt_0) \Rightarrow S_\infty$. Since $W(0, \vt_0)$ is a deterministic vector, $\nproc{2}(0) = S_n^T W(0, \vt_0) + r_n(0)$ converges as well and is thus tight in $\R$ by Prokhorov's Theorem. It remains to show that for every $\eps, \eta > 0$, there exists a $\delta > 0$ and $n_0 \in \N$ such that for all $n \ge n_0$
	\begin{align}
		\label{eq:tight_c}
		\prob{\sup_{\abs{s-t}<\delta} \abs{\nproc{2}(s) - \nproc{2}(t)} \ge \eps} \le \eta.
	\end{align}
	For any $\eps, \eta > 0$, we have
	\begin{align*}
		&\prob{\sup_{\abs{s-t}<\delta} \abs{\nproc{2}(s) - \nproc{2}(t)} \ge \eps}\\
		&\quad= \prob{\sup_{\abs{s-t}<\delta} \abs{S_n^T \Big(W(s,\vt_0) - W(t,\vt_0)\Big) + r^{(2)}_n(s) - r^{(2)}_n(t)} \ge \eps}\\
		&\quad\le \prob{\norm{S_n} \sup_{\abs{s-t}<\delta} \norm{\left(W(s,\vt_0) - W(t,\vt_0)\right)} \ge \frac{\eps}{3}} + 2 \cdot \prob{\sup_{t \in \R} \abs{r^{(2)}_n(t)} \ge \frac{\eps}{3}}\\
		&\quad= \prob{\norm{S_n} \ge \frac{\eps}{3 \cdot \sup_{\abs{s-t}<\delta} \norm{\left(W(s,\vt_0) - W(t,\vt_0)\right)}}} + 2 \cdot \prob{\sup_{t \in \R} \abs{r^{(2)}_n(t)} \ge \frac{\eps}{3}}.
	\end{align*}
	The second summand clearly converges to zero as $n$ goes to infinity. For the first summand, choose $c_\infty$ such that $\prob{\norm{S_\infty} \ge c_\infty} \le \frac{\eta}{2}$ and $n_0 \in \N$ such that for all $n \ge n_0$
	\begin{align*}
		\prob{\norm{S_n} \ge c_\infty} &\le \abs{\prob{\norm{S_n} \ge c_\infty} - \prob{\norm{S_\infty} \ge c_\infty}} + \prob{\norm{S_\infty} \ge c_\infty}\\
		&\le \frac{\eta}{2} + \frac{\eta}{2} = \eta.
	\end{align*}
	Since $W(t, \vt_0)$ is uniformly continuous in $t$, we can find a $\delta > 0$ such that
	\begin{align*}
		\frac{\eps}{3 \cdot \sup_{\abs{s-t}<\delta} \norm{\left(W(s,\vt_0) - W(t,\vt_0)\right)}} \ge c_\infty,
	\end{align*}
	which concludes the proof of (\ref{eq:tight_c}). Note that uniform tightness in $\cspace$ implies uniform tightness in the larger space $\lspace$, which in turn implies asymptotic tightness in $\lspace$.
	
	Due to their asymptotic linear representations, the fidis of $\nproc{1}$ and $\nproc{2}$ converge to a centered normal distribution by the multivariate CLT. Having established their asymptotic tightness as well, we can apply \citet[Theorem 7.17]{kosorok2008introduction} to follow that both processes and thus also their sum $\tnproc$ converge weakly to a centered Gaussian process in $\lspace$. The only step left to prove the statement of the theorem is the calculation of the covariance structure of the limiting process. The auto-covariance functions of $\nproc{1}(t)$ and $\nproc{2}(t)$ are already given above. For their cross-covariance function, we get
	\begin{align*}
		&\cov[\vt_0]{\xi^{(1)}(s,X,Y), \xi^{(2)}(t,X,Y)}\\
		&\quad= \expec[\vt_0]{\qty(\ind{Y \le s} - F(s|\vt_0,X))\qty(-L^T(X,Y;\vt_0) W(t,\vt_0))}\\
		&\quad= W^T(t,\vt_0) \expec[\vt_0]{\qty(F(s|\vt_0,X) - \ind{Y \le s}) L(X,Y;\vt_0)}. \qedhere
	\end{align*} 
\end{proof}

\begin{proof}[Proof of \fref{cor:conv_proc}]
	We will show that \fref{ass:m1p} implies \ref{ass:m1}. Note that \ref{ass:m1}\fref{cond:W_unif_cont} matches \ref{ass:m1p}\fref{cond:W_cont}, so we only need to show that \fref{ass:m1}\fref{cond:F_diff} and \fref{cond:w_unif_wlln} hold under \ref{ass:m1p}. 
		
		By \fref{ass:m1p}\fref{cond:v_ddom}, $\vt \mapsto f(t|\vt, x)$ is differentiable on $V$, and we can apply Lebesgue's Dominated Convergence Theorem (DCT) to deduce that
	\begin{align*}
		\frac{\del}{\del \vt} F(t|\vt, X_i) = \frac{\del}{\del \vt} \int_{-\infty}^t f(s|\vt, X_i) \nu(ds) = \int_{-\infty}^t \frac{\del}{\del \vt} f(s|\vt, X_i) \nu(ds),
	\end{align*}
	which implies that $\vt \mapsto F(t|\vt, x)$ is differentiable on $V$.
	
	Let $\eps > 0$ and $\{r_n\}_{n \ge 1}$ be a non-negative sequence with $r_n \to 0$ as $n \to \infty$. By Markov's inequality, we have
	\begin{align}
		&\prob{\sup_{t \in \bar{\R}} \sup_{\norm{\vt-\vt_0}\le r_n} \norm{\frac{1}{n} \sumin w(t, \vt, X_i) - W(t, \vt_0)} > \eps} \nonumber\\
		&\quad\le \prob{\frac{1}{n} \sumin \sup_{t \in \bar{\R}} \sup_{\norm{\vt-\vt_0}\le r_n} \norm{w(t, \vt, X_i) - w(t, \vt_0, X_i)} > \frac{\eps}{2}} \nonumber\\
		&\qquad + \prob{\sup_{t \in \bar{\R}} \norm{\frac{1}{n} \sumin w(t, \vt_0, X_i) - W(t, \vt_0)} > \frac{\eps}{2}} \nonumber\\
		&\quad \le \frac{2}{\eps} \cdot \expec{\sup_{t \in \bar{\R}} \sup_{\norm{\vt-\vt_0} \le r_n} \norm{w(t, \vt, X) - w(t, \vt_0, X)}} \nonumber\\
		&\qquad + \prob{\sup_{t \in \bar{\R}} \norm{\frac{1}{n} \sumin w(t, \vt_0, X_i) - W(t, \vt_0)} > \frac{\eps}{2}}. \label{eq:w_wlln}
	\end{align}
	By \fref{ass:m1p}\fref{cond:w_cont}, the integrand of the first summand goes to zero as $n$ increases. Due to \fref{ass:m1p}\fref{cond:v_ddom}, we can apply Lebesgue's DCT to follow that the expectation converges to zero as well. For that, note that because of \fref{ass:m1p}\fref{cond:v_ex_dom}, the order of integration and differentiation can be interchanged, and we have 
	\begin{align*}
		\norm{w(t,\vt,x)} &= \norm{\int_{-\infty}^t v(s,\vt,x) \nu(ds)}\\
		&\le \int_{-\infty}^t \norm{v(s,\vt,x)} \nu(ds)\\
		&\le \sup_{\tilde{\vt} \in V} \int \norm{v(s,\tilde{\vt},x)} \nu(ds),
	\end{align*}
	which is integrable according to \fref{ass:m1p}\fref{cond:v_ddom}.
	Turning to the second summand on the right-hand side of \fref{eq:w_wlln}, we have for any $K > 0$
	\begin{align*}
		&\sup_{t \in \bar{\R}} \norm{\frac{1}{n} \sumin w(t,\vt_0,X_i) - W(t,\vt_0)}\\
		&\quad= \sup_{t \in \bar{\R}} \norm{\int w(t,\vt_0,x) (H_n - H)(dx)}\\
		&\quad= \sup_{t \in \bar{\R}} \norm{\int \int_{-\infty}^t v(s,\vt_0,x) \nu(ds) (H_n-H)(dx)}\\
		&\quad\le \sup_{|t| \le K} \norm{\int \int_{-K}^t v(s,\vt_0,x) \nu(ds) (H_n-H)(dx)}\\
		&\qquad+ \int \int \ind{|s|>K} \norm{v(s,\vt_0,x)} \nu(ds) H_n(dx) + \int \int \ind{|s|>K} \norm{v(s,\vt_0,x)} \nu(ds) H(dx)\\
		&\quad\eqqcolon A_{1,n}(K) + A_{2,n}(K) + A_3(K).
	\end{align*}
	By \citet[Theorem 2]{jennrich1969asymptotic}, applicable due to \fref{ass:m1p}, $\lim_{n \to \infty} A_{1,n}(K) = 0$ for any value of $K$. Moreover, an iterated application of Lebesgue's DCT using \fref{ass:m1p} to find dominating integrable functions, yields
	\begin{align*}
		\lim_{K \to \infty} A_3(K) = \int \int \lim_{K \to \infty} \ind{|s|>K} \norm{v(s,\vt_0,x)} \nu(ds) H(dx) = 0.
	\end{align*}
	Since, by SLLN, $\lim_{n \to \infty} A_{2,n}(K) = A_3(K)$ wp1, the second term also vanishes as $n$ and $K$ tend to infinity. Altogether, it follows that the right-hand side of \fref{eq:w_wlln} converges to zero.
\end{proof}

\pagebreak
\begin{proof}[Proof of \fref{thm:conv_proc_boot}]		
	The proof is similar to the proof of convergence of the original test statistic. We start by splitting the process as follows
	\begin{align*}
		\tsnproc(t) &= \underbrace{\frac{1}{\sqrt{n}} \sumin \qty(\textcolor{black}{\ind{\Ys \le t} - F(t|\vtn,X_i)})}_{\nproc{1*}(t)}\\
		&\quad+ \underbrace{\frac{1}{\sqrt{n}} \sumin \qty(\textcolor{black}{F(t|\vtn,X_i) - F(t|\vtns,X_i)})}_{\nproc{2*}(t)}.
	\end{align*}
	To prove the convergence of $\tsnproc$, we will again apply \citet[Theorem 7.17]{kosorok2008introduction}. We start by proving the convergence of $\nproc{1*}$ using \citet[Theorem 11.16]{kosorok2008introduction}. Since the $X_i$ are fixed under $\P_n^*$, we can write
	\begin{align*}
		\nproc{1*}(t) &= \frac{1}{\sqrt{n}} \sumin \qty(\textcolor{black}{\ind{\Ys \le t} - F(t|\vtn,X_i)})\\
		&= \sumin \qty(\textcolor{black}{\tilde{f}_{n,i}(t) - \expecs{\tilde{f}_{n,i}(t)}}),
	\end{align*}
	where $\tilde{f}_{n,i}(t) = \frac{1}{\sqrt{n}} \ind{\Ys \le t}$ with envelope $\tilde{F}_{n,i} = \frac{1}{\sqrt{n}}$. The separability of $\{\tilde{f}_{n,i}\}$ can be shown with a similar argument as in the original convergence proof using the fact that $\tilde{f}_{n,i}(t)$ is right-continuous. By \citet[Lemma 11.15]{kosorok2008introduction}, it follows that the triangular array is almost measurable Suslin. The manageability condition (A) of \citet[Theorem 11.16]{kosorok2008introduction} is fulfilled since the indicator functions in $\tilde{f}_{n,i}(t)$ are monotone increasing in $t$ (see page 221 in Kosorok's book). The second condition (B) holds as
	\begin{align*}
		\tilde{K}(s,t) &\coloneqq \lim_{n \to \infty} \expecs{\nproc{1*}(s) \nproc{1*}(t)}\\
		&= \lim_{n \to \infty} \frac{1}{n} \expecs{\sumin \sum_{j=1}^n \left(\ind{\Ys \le s} - F(s|\vtn,X_i)\right) \left(\ind{Y_{j,n}^* \le t} - F(t|\vtn,X_j)\right)}\\
		&=_{(1)} \lim_{n \to \infty} \frac{1}{n} \sumin \expecs{\left(\ind{\Ys \le s} - F(s|\vtn,X_i)\right) \left(\ind{\Ys \le t} - F(t|\vtn,X_i)\right)}\\
		&=_{(2)} \expec[\vt_0]{\left(\ind{Y \le s} - F(s|\vt_0,X)\right) \left(\ind{Y \le t} - F(t|\vt_0,X)\right)}\\
		&= \prob[\vt_0]{Y \le \min(s,t)} - \expec{F(s|\vt_0,X) F(t|\vt_0,X)} < \infty,
	\end{align*}
	where for the equality $=_{(1)}$, we used the fact that $Y_{1,n}^*, \dots, Y_{n,n}^*$ are independent, and in $=_{(2)}$, we used \fref{ass:me2}\fref{cond:conv_I-F}. The verification of the next two conditions, (C) and (D), is straightforward:
	\begin{align*}
		\limsup_{n \to \infty} \sumin \expecs{\tilde{F}_{n,i}^2} = \limsup_{n \to \infty} \sumin \frac{1}{n} = 1 < \infty
	\end{align*}
	and
	\begin{align*}
		\sumin \expecs{\tilde{F}_{n,i} \ind{\tilde{F}_{n,i} > \eps}} = \sumin \expecs{\frac{1}{n} \ind{\frac{1}{\sqrt{n}} > \eps}} = \ind{\frac{1}{\sqrt{n}} > \eps} \xrightarrow[n \to \infty]{} 0 \quad \forall \eps > 0.
	\end{align*}
	For condition (E), we need to consider $\rho_n(s,t) \coloneqq \qty(\sumin \expecs{\abs{\tilde{f}_{n,i}(s) - \tilde{f}_{n,i}(t)}^2})^{\frac{1}{2}}$. By \fref{ass:me2}\fref{cond:conv_I}, we have
	\begin{align*}
		\rho_n(s,t) %
		&= \qty(\frac{1}{n} \sumin \expecs{\abs{\ind{\Ys \le s} - \ind{\Ys \le t}}})^{\frac{1}{2}}\\
		&\xrightarrow[n \to \infty]{} \qty(\expec[\vt_0]{\abs{\ind{Y \le s} - \ind{Y \le t}}})^{\frac{1}{2}} \eqqcolon \rho(s,t).
	\end{align*}
	Assume that $\{s_n\}, \{t_n\} \in \R$ with $\rho(s_n, t_n) \to 0$. Then we have
	\begin{align*}
		\rho_n(s_n,t_n) %
		&\le \sup_{s,t \in \bar{\R}} \abs{\rho_n(s,t) - \rho(s,t)} + \rho(s_n, t_n).
	\end{align*}
	The second summand converges to zero by assumption. Since, similarly to the reverse triangle inequality, it holds $\sqrt{|a-b|} \ge \abs{\sqrt{a} - \sqrt{b}}$, we get for the first summand
	\begin{align*}
		&\sup_{s,t \in \bar{\R}} \abs{\rho_n(s,t) - \rho(s,t)}^2\\
		&\quad= \sup_{s,t \in \bar{\R}} \abs{\qty(\frac{1}{n} \sumin \expecs{\abs{\ind{\Ys \le s} - \ind{\Ys \le t}}})^{\frac{1}{2}} - \qty(\expec[\vt_0]{\abs{\ind{Y \le s} - \ind{Y \le t}}})^{\frac{1}{2}}}^2\\
		&\quad\le \sup_{s,t \in \bar{\R}} \abs{\frac{1}{n} \sumin \expecs{\abs{\ind{\Ys \le s} - \ind{\Ys \le t}}} - \expec[\vt_0]{\abs{\ind{Y \le s} - \ind{Y \le t}}}}\\
		&\quad\le \sup_{s,t \in \bar{\R}} \abs{\frac{1}{n} \sumin \expecs{\ind{\Ys \le s} + \ind{\Ys \le t}} - \expec[\vt_0]{\ind{Y \le s} + \ind{Y \le t}}}\\
		&\quad\le 2 \sup_{t \in \bar{\R}} \abs{\frac{1}{n} \sumin \expecs{\ind{\Ys \le t}} - \expec[\vt_0]{\ind{Y \le t}}}\\
		&\quad= 2 \sup_{t \in \bar{\R}} \abs{\expec{F(t|\vt_0,X)} - \frac{1}{n} \sumin F(t|\vtn, X_i)}\\
		&\quad\le 2 \left(\sup_{t \in \bar{\R}} \abs{\expec{F(t|\vt_0,X)} - \frac{1}{n} \sumin F(t|\vt_0, X_i)} \right.\\
		&\qquad\qquad \left. + \sup_{t \in \bar{\R}} \abs{\frac{1}{n} \sumin \qty(\textcolor{black}{F(t|\vt_0, X_i) - F(t|\vtn, X_i)})}\right).
	\end{align*}
	The almost sure convergence of the first summand to zero is equivalent to saying that $\tilde{\F} = \{x \mapsto F(t|\vt_0,x) ~|~ t \in \R\}$ is a Glivenko-Cantelli class. This is in fact the case as we showed $\tilde{\F}$ to be Donsker in the proof of \fref{thm:conv_proc} and every Donsker class is also Glivenko-Cantelli (see \citet[Lemma 8.17]{kosorok2008introduction}). To analyze the convergence of the second summand, we use the mean value theorem, Cauchy-Schwarz and triangle inequality, yielding
	\begin{align}
		&\sup_{t \in \bar{\R}} \abs{\frac{1}{n} \sumin \qty(\textcolor{black}{F(t|\vt_0, X_i) - F(t|\vtn, X_i)})} \nonumber\\ 
		&\quad= \sup_{t \in \bar{\R}} \abs{\qty(\vtn - \vt_0)^T \frac{1}{n} \sumin w(t, \tilde{\vt}_n(t, X_i), X_i)} \nonumber\\
		&\quad\le \norm{\vtn - \vt_0} \sup_{t \in \bar{\R}} \norm{\frac{1}{n} \sumin w(t, \tilde{\vt}_n(t, X_i), X_i)} \nonumber\\
		&\quad\le \norm{\vtn - \vt_0} \Big(\underbrace{\sup_{t \in \bar{\R}} \norm{\frac{1}{n} \sumin w(t, \tilde{\vt}_n(t, X_i), X_i) - W(t, \vt_0)}}_{R_n} + \sup_{t \in \bar{\R}} \norm{W(t, \vt_0)}\Big),
		\label{eq:cond_E_summand2}
	\end{align}
	where $\tilde{\vt}_n(t, X_i)$ lies on the line segment connecting $\vt_0$ and $\vtn$ and may depend on $t$ and $X_i$. By assumption, $\norm{\vtn - \vt_0}$ converges to zero wp1. Moreover, $\sup_{t \in \bar{\R}} \norm{W(t, \vt_0)}$ is bounded according to \fref{ass:m2}\fref{cond:W_unif_bdd}. So the desired result, namely almost sure convergence of \fref{eq:cond_E_summand2} to zero, follows if $R_n$ is appropriately bounded. Since $\vtn \to \vt_0$ in pr.\ by \fref{ass:e1}\fref{cond:mle_lin}, there exists a non-negative sequence $\{r_n\}_{n \ge 1}$ such that $r_n \to 0$ and $\prob{\norm{\vtn - \vt_0} > r_n} \to 0$. This implies that
	\begin{align*}
		\prob{R_n \le \sup_{t \in \bar{\R}} \sup_{\norm{\vt - \vt_0} \le r_n} \norm{\frac{1}{n} \sumin w(t, \vt, X_i) - W(t, \vt_0)}} \to 1.
	\end{align*}
	By \fref{ass:m2}\fref{cond:w_unif_slln}, it follows that $R_n \to 0$ wp1.
	Hence, the product on the right-hand side of (\ref{eq:cond_E_summand2}) converges to zero wp1. In summary, we have verified all five conditions (A)-(E) of \citet[Theorem 11.16]{kosorok2008introduction} and thus can conclude that $\nproc{1*}$ converges to a tight centered Gaussian process with auto-covariance function $\tilde{K}(s,t)$. By \citet[Lemma 7.12]{kosorok2008introduction}, it follows that $\nproc{1*}$ is asymptotically tight.
	\pagebreak
	
	Next, we investigate $\nproc{2*}$. In particular, we will find an asymptotically equivalent representation. Using a Taylor expansion and \fref{ass:e2}\fref{cond:mle_lin_boot}, we have
	\begin{align*}
		\nproc{2*}(t) &= \frac{1}{\sqrt{n}} \sumin \qty(\textcolor{black}{F(t|\vtn, X_i) - F(t|\vtns, X_i)})\\
		&= -\sqrt{n} \qty(\vtns - \vtn)^T \frac{1}{n} \sumin \frac{\del}{\del \vt} F(t|\tilde{\vt}_n^*(t,X_i), X_i)\\
		&= - \qty(n^{-1/2} \sumin L^T(X_i, \Ys, \vtn) + \opIs) \frac{1}{n} \sumin w(t, \tilde{\vt}_n^*(t,X_i), X_i),
	\end{align*}
	where $\tilde{\vt}_n^*(t,X_i)$ lies on the line segment connecting $\vtn$ and $\vtns$ and may depend on $t$ and $X_i$.
	From Assumptions \ref{ass:e1}\fref{cond:mle_lin} and \ref{ass:e2}\fref{cond:mle_lin_boot} together with the WLLN, it follows that $\vtns$ converges to $\vt_0$ in pr.\ and thus there exists a non-negative sequence $\{r_n\}_{n \ge 1}$ such that $r_n \to 0$ and $\prob{\norm{\vtns - \vt_0} > r_n} \to 0$ as $n \to \infty$. Using \fref{ass:m1}\fref{cond:w_unif_wlln} and a similar argument as in the proof of \fref{thm:conv_proc}, this further yields
	\begin{align*}
	\sup_{t \in \bar{\R}} \norm{\frac{1}{n} \sumin w(t,\tilde{\vt}_n^*(t,X_i), X_i) - W(t,\vt_0)} = \opIs.
	\end{align*}
	Altogether, we get wp1
	\begin{align*}
		\nproc{2*}(t) = n^{-1/2} \sumin -W^T(t,\vt_0) L(X_i, \Ys, \vtn) + \opIs
	\end{align*}
	uniformly in $t$. Since the sum on the right-hand side is asymptotically equivalent to $\nproc{2*}(t)$ in the sense of \citet[Lemma 7.23(i)]{kosorok2008introduction}, we will substitute it from now on without mentioning it again.
	
	To analyze the convergence of the fidis of $\tsnproc$, we use Cram\'er-Wold device (see e.g. \citet[Theorem 7.7]{billingsley1968convergence}). So we need to show that 
	\begin{align}
		\label{eq:cramwol}
		\forall k \in \N, ~(t_1,\dots,t_k) \in \R^k, ~0 \ne a \in \R^k:\quad \sumjk a_j \tsnproc(t_j) \Rightarrow \mathcal{N}(0, a^t\Sigma a) \quad \text{wp1},
	\end{align}
	where $\Sigma_{j,l} = K(t_j, t_l)$ for $1 \le j,l \le k$. We have
	\begin{align*}
		Z_n^* &= \sumjk a_j \tsnproc(t_j)\\
		&= \sumin \underbrace{\sumjk a_j \frac{1}{\sqrt{n}} \qty(\ind{\Ys \le t_j} - F(t_j|\vtn, X_i))}_{\xi_{i,n}^*} - \underbrace{\sumjk a_j W^T(t_j, \vt_0)}_{A^T} \underbrace{\frac{1}{\sqrt{n}} L(X_i, \Ys, \vtn)}_{\eta_{i,n}^{*}}
	\end{align*}
	with $\expecs{\xi_{i,n}^*} = \expecs{\eta_{i,n}^*} = 0$ and $(\xi_{1,n}^*, \eta_{1,n}^*), \dots, (\xi_{n,n}^*, \eta_{n,n}^*)$ independent. It follows that
	\begin{align*}
		\vars{Z_n^*} &= \sumin \vars{\xi_{i,n}^*} + \vars{A^T \eta_{i,n}^*} 	- 2\covs{\xi_{i,n}^*, A^T \eta_{i,n}^*}\\
		&= \sumin \expecs{\xi_{i,n}^{*2}} + A^T \expecs{\eta_{i,n}^* \eta_{i,n}^{*T}} A - 2A^T \expecs{\xi_{i,n}^* \eta_{i,n}^*}.
	\end{align*}
	For the first summand, we get
	\begin{align*}
		&\sumin \expecs{\xi_{i,n}^{*2}} \\
		&\quad= \frac{1}{n} \sumin \sumjk 	\sum_{l=1}^k a_j a_l \expecs{\qty(\ind{\Ys \le t_j} - F(t_j|\vtn, X_i))\qty(\ind{\Ys \le t_l} - F(t_l|\vtn, X_i))}\\
		&\quad= \sumjk \sum_{l=1}^k a_j \qty( \frac{1}{n} \sumin \expecs{\ind{\Ys \le \min(t_j,t_l)}} - F(t_j|\vtn,X_i)F(t_l|\vtn,X_i) ) a_l\\
		&\quad\xrightarrow[n \to \infty]{} \sumjk \sum_{l=1}^k a_j \Big( \expec[\vt_0]{\ind{Y \le \min(t_j,t_l)}} - \expec{F(t_j|\vt_0,X) F(t_l|\vt_0,X)} \Big) a_l\\
		&\quad= \sumjk \sum_{l=1}^k a_j \tilde{K}(t_j,t_l) a_l,
	\end{align*}
	using \fref{ass:me2}\fref{cond:conv_I-F}. By \fref{ass:me2}\fref{cond:conv_LL}, the second summand converges as well:
	\begin{align*}
		\sumin A^T \expecs{\eta_{i,n}^* \eta_{i,n}^{*T}} A &= A^T \frac{1}{n} \sumin \expecs{L(X_i,\Ys,\vtn) L^T(X_i,\Ys,\vtn)} A\\
		&\xrightarrow[n \to \infty]{} A^T \expec[\vt_0]{L(X,Y,\vt_0) L^T(X,Y,\vt_0)} A\\
		&= \sumjk \sum_{l=1}^k a_j W^T(t_j, \vt_0) 	\cov[\vt_0]{L(X,Y,\vt_0)} W(t_l,\vt_0) a_l.
	\end{align*}
	Finally, an application of \fref{ass:me2}\fref{cond:conv_(I-F)L} yields
	\begin{align*}
		&\sumin A^T \expecs{\xi_{i,n}^* \eta_{i,n}^*}\\
		&\quad= \sumjk \sum_{l=1}^k a_j W^T(t_j,\vt_0) \frac{1}{n} \sumin \expecs{\qty(\ind{\Ys \le t_l} - F(t_l|\vtn, X_i)) L(X_i, \Ys, \vtn)} a_l\\
		&\quad\xrightarrow[n \to \infty]{} \sumjk \sum_{l=1}^k a_j 	W^T(t_j,\vt_0) \expec[\vt_0]{\qty(\ind{Y \le t_l} - F(t_l|\vt_0,X)) L(X,Y,\vt_0)} a_l.
	\end{align*}
	In summary, we have shown that $\vars{Z_n^*} \to a^t \Sigma a$ wp1. Since $\Sigma$ is a covariance matrix and hence positive semidefinite, we know that $a^t \Sigma a \ge 0$. If $a^T \Sigma a = 0$, $Z_n^* \Rightarrow \mathcal{N}(0,a^T \Sigma a) \equiv \mathcal{N}(0,0)$ trivially holds. Otherwise, i.e.\ if $a^T \Sigma a > 0$, we can apply \citet[Corollary to Theorem 1.9.3]{serfling2009approximation} and thus have to verify the Lyapunov condition:
	\begin{align*}
		\frac{1}{\vars{Z_n^*}^{(2+v)/2}} \sumin \expecs{\abs{\xi_{i,n}^* - 	A^T \eta_{i,n}^*}^{2+v}} \to 0 \quad \text{wp1}
	\end{align*}
	for some $v>0$, where the null set does not depend on $a$. Since $\vars{Z_n^*}$ converges to $a^T\Sigma a > 0$, it is sufficient to prove that there exists $v>0$ such that the sum of expectations converges to zero wp1. Note that
	\begin{align*}
		\sumin \expecs{\abs{\xi_{i,n}^* - A^T \eta_{i,n}^*}^{2+v}} %
		&\le \sumin \qty(\expecs{\abs{\xi_{i,n}^*}^{2+v}}^{1/(2+v)} + \expecs{\abs{A^T \eta_{i,n}^*}^{2+v}}^{1/(2+v)})^{2+v}\\
		&\le 2^{2+v} \qty(\sumin \expecs{\abs{\xi_{i,n}^*}^{2+v}} + \sumin \expecs{\abs{A^T \eta_{i,n}^*}^{2+v}}),
	\end{align*}
	so we can analyze the two sums separately. We have
	\begin{align*}
		\sumin \expecs{\abs{\xi_{i,n}^*}^{2+v}} &= \sumin \expecs{\abs{\frac{1}{\sqrt{n}} \sumjk a_j \qty(\ind{\Ys \le t_j} - F(t_j|\vtn, X_i))}^{2+v}}\\
		&\le \frac{1}{n^{v/2}} \frac{1}{n} \sumin \expecs{\qty(\sumjk \abs{a_j})^{2+v}}\\
		&= \frac{1}{n^{v/2}} \qty(\sumjk \abs{a_j})^{2+v} 
		\xrightarrow[n \to \infty]{} 0,
	\end{align*}
	and using Cauchy-Schwarz inequality
	\begin{align*}
		\sumin \expecs{\abs{A^T \eta_{i,n}^*}^{2+v}} &= \sumin 	\expecs{\abs{\frac{1}{\sqrt{n}} \sumjk a_j W^T(t_j, \vt_0) L(X_i, \Ys, \vtn)}^{2+v}}\\
		&\le \frac{1}{n^{v/2}} \norm{\sumjk a_j W(t_j, \vt_0)}^{2+v} \frac{1}{n} \sumin \expecs{\norm{L(X_i, \Ys, \vtn)}^{2+v}}.
	\end{align*}
	As $n^{-v/2}$ converges to zero as $n$ goes to infinity, it remains to show that the empirical mean on the right-hand side is bounded. For that, choose $v=\delta$ from \fref{ass:e2}\fref{cond:L_delta} and note that for $n$ large enough $\vtn \in V$ as $\vtn \to \vt_0$ wp1, such that
	\begin{align*}
		\expecs{\norm{L(x, \Ys, \vtn)}^{2+\delta}} &= \int \norm{L(x, y, \vtn)}^{2+\delta} f(y|\vtn,x) \nu(dy)\\
		&\le \sup_{\vt \in V} \int \norm{L(x, y, \vt)}^{2+\delta} f(y|\vt,x) \nu(dy).
	\end{align*} 
	By \fref{ass:e2}\fref{cond:L_delta}, we can use the SLLN to conclude that wp1
	\begin{align*}
		&\frac{1}{n} \sumin \expecs{\norm{L(X_i, \Ys, \vtn)}^{2+\delta}} \le \frac{1}{n} \sumin \sup_{\vt \in V} \int \norm{L(X_i, y, \vt)}^{2+\delta} f(y|\vt,X_i) \nu(dy)\\
		&\quad \xrightarrow[n \to \infty]{} \int \sup_{\vt \in V} \int \norm{L(x, y, \vt)}^{2+\delta} f(y|\vt,x) \nu(dy) H(dx) < \infty.
	\end{align*}
	Therefore, the Lyapunov condition is fulfilled and we can follow that (\ref{eq:cramwol}) holds and the fidis of the process $\tsnproc$ converge to a multivariate normal distribution.
	
	Following \citet[Theorem 7.17]{kosorok2008introduction}, the proof is complete if we can show that the process is asymptotically tight in $\lspace$. As already mentioned, the asymptotic tightness of $\nproc{1*}$ follows by \citet[Lemma 7.12]{kosorok2008introduction} from its convergence to a tight process. The proof of \citet[Lemma 6.30]{dikta2021bootstrap} shows that $n^{-1/2} \sumin L(X_i, \Ys, \vtn)$ converges to a zero mean multivariate normal distribution. This allows us to apply the same arguments to $\nproc{2*}$ as used in \fref{thm:conv_proc} to verify the asymptotic tightness of $\nproc{2}$.
\end{proof}

\begin{proof}[Proof of \fref{cor:conv_proc_boot}]
	We will first show that \fref{ass:m1p} implies \ref{ass:m2}\fref{cond:W_unif_bdd} and that \ref{ass:me2p} implies \ref{ass:me2}\fref{cond:conv_I}, \fref{cond:conv_I-F} and \fref{cond:conv_(I-F)L}. In a third step, we will illustrate how the proof of \fref{thm:conv_proc_boot} can be modified such that \fref{ass:m2}\fref{cond:w_unif_slln} is not needed, given that \ref{ass:m1p} holds.
	
	By \fref{ass:m1p}\fref{cond:v_ex_dom}, we have $w(t, \vt_0, x) = \int_{-\infty}^t v(s, \vt_0, x) \nu(ds)$ and hence
	\begin{align*}
		\sup_{t \in \bar{\R}} \norm{W(t, \vt_0)} &= \sup_{t \in \bar{\R}} \norm{\int \int_{-\infty}^t v(s, \vt_0, x) \nu(ds) H(dx)}\\
		&\le \int \int \norm{v(s,\vt_0,x)} \nu(ds) H(dx),
	\end{align*}
	which is finite according to \fref{ass:m1p}\fref{cond:v_ddom}.
	
	We can use \citet[Lemma 5.58]{dikta2021bootstrap} to show that the convergence assumptions in \ref{ass:me2}\fref{cond:conv_I}, \fref{cond:conv_I-F} and \fref{cond:conv_(I-F)L} hold under \fref{ass:me2p}. Note that in \fref{thm:conv_proc_boot}, we assume that $\vtn \to \vt_0$ wp1 and the density $f$ is continuous in $\vt$ at $\vt_0$. So, in order for \citet[Lemma 5.58]{dikta2021bootstrap} to be applicable, we only need to ensure that the functions $\ell_1, \ell_2$ and $\ell_3$ given in \ref{ass:me2}\fref{cond:conv_I}, \fref{cond:conv_I-F} and \fref{cond:conv_(I-F)L} are continuous and that there exist open neighborhoods $V_1$ and $V_2$ of $\vt_0$ such that
	\begin{align}
		\label{eq:cond_5.58}
		\int \int \sup_{\vt_1 \in V_1} \abs{\ell_k(x, y, \vt_1)} \sup_{\vt_2 \in V_2} f(y|\vt_2, x) \nu(dy) H(dx) < \infty.
	\end{align}
	Observe that $\ell_1$ is independent of $\vt$, $\ell_2$ is continuous in $\vt$ since $F$ is assumed to be continuous in $\vt$ and $\ell_4$ is continuous in $\vt$ for the same reason and by \fref{ass:me2p}\fref{cond:xi_cont_int}. It is easy to see that $\ell_1$ and $\ell_2$ are absolutely bounded above by $1$ and $\abs{\ell_4(x,y,\vt)} \le \abs{L(x,y,\vt)}$, so the validity of \fref{eq:cond_5.58} immediately follows from \fref{ass:me2p}.
	
	In the proof of \fref{thm:conv_proc_boot}, \fref{ass:m2}\fref{cond:w_unif_slln} is only used once, namely in order to show that (\ref{eq:cond_E_summand2}) converges to zero wp1. Here, we will prove this assertion by establishing that $\sup_{t \in \bar{\R}} \norm{\frac{1}{n} \sumin w(t, \tilde{\vt}(t, X_i), X_i)}$ is asymptotically bounded wp1, where $\tilde{\vt}(t, X_i)$ lies on the line segment connecting $\vt_0$ and $\vtn$.
	Since $\vtn$ converges almost surely to $\vt_0$, $\tilde{\vt}_n(t,X_i)$ will eventually lie in $V$. It follows that, for sufficiently large $n$,
	\begin{align*}
		\sup_{t \in \R} \norm{\frac{1}{n} \sumin w(t, \tilde{\vt}_n(t, X_i), X_i)}
		&\le \sup_{\vt \in V} \sup_{t \in \R} \norm{\frac{1}{n} \sumin w(t, \vt, X_i)}\\
		&\le \frac{1}{n} \sumin \sup_{\vt \in V} \sup_{t \in \R} \norm{w(t,\vt,X_i)}\\
		&\le \frac{1}{n} \sumin \sup_{\vt \in V} \int \norm{v(s,\vt,X_i)} \nu(ds)\\
		&\eqqcolon \frac{1}{n} \sumin M_v(X_i).
	\end{align*}
	According to \fref{ass:m1p}\fref{cond:v_ddom}, $\expec{M_v(X)}$ is finite. Thus, by the SLLN, the arithmetic mean converges wp1 to the finite expected value.
\end{proof}

\begin{proof}[Proof of \fref{thm:cons}]
	With $F_1$ denoting the true conditional distribution function underlying the sample $\{(X_i, Y_i)\}_{i=1}^n$, we can write
	\begin{align*}
		\frac{1}{\sqrt{n}} \norm{\tnproc} &= \sup_{t \in \R} \abs{\frac{1}{n} \sumin \qty(\textcolor{black}{\ind{Y_i \le t} - F(t|\vtn, X_i)})}\\
		&= \sup_{t \in \R} \Bigg| \underbrace{\frac{1}{n} \sumin \ind{Y_i \le t} - \expec{F_1(t|X)}}_{T_{1,n}(t)} + \underbrace{\expec{F_1(t|X) - 	F(t|\vt_1,X)}}_{T_2(t)} \\
		&\quad\qquad+ \underbrace{\expec{F(t|\vt_1, X)} - \frac{1}{n} 	\sumin F(t|\vt_1, X_i)}_{T_{3,n}(t)}\\
		&\quad\qquad+ \underbrace{\frac{1}{n} \sumin \qty(\textcolor{black}{F(t|\vt_1, X_i) - F(t|\vtn, X_i)})}_{T_{4,n}(t)} \Bigg|.
	\end{align*}
	The classical Glivenko-Cantelli Theorem states that $\sup_{t \in \R} \abs{T_{1,n}(t)} \to 0$ a.s. A similar result for $T_{3,n}$ can be proven using generalized empirical process theory and the fact that $\tilde{\F} = \{F(t|\vt_1, \cdot) ~|~ t \in \R\}$ is a Donsker class as illustrated in the proof of \fref{thm:conv_proc}. Using a Taylor expansion and arguments along the line of the proof of \fref{thm:conv_proc}, it can be shown that $\sup_{t \in \R} \abs{T_{4,n}(t)}$ also converges to zero in pr. As a consequence, we can write
	\begin{align*}
		\frac{1}{\sqrt{n}} \norm{\tnproc} = \sup_{t \in \R} \abs{T_2(t)} + \opI.
	\end{align*}
	
	By \fref{ass:h1}, there exists a $t \in \R$ such that $\abs{T_2(t)} > 0$ and hence \mbox{$\sup_{t \in \R} \abs{T_2} > \frac{1}{k}$} for some $k \ge 1$.
	Since, by \fref{thm:conv_proc_boot} and the Continuous Mapping Theorem, the bootstrap test statistic $\norm{\tsnproc}$ converges in distribution to $\norm{\tinfproc}$ (even under $H_1$), the sequence of bootstrap critical values $c_n$ converges to a constant. It follows that (for \mbox{sufficiently large $n$)}
	\begin{equation*}
		\prob{\norm{\tnproc} > c_n} \ge \prob{\frac{1}{\sqrt{n}} \norm{\tnproc} > \frac{1}{k}} = \prob{\sup_{t \in \R} \abs{T_2} + \opI > \frac{1}{k}} \xrightarrow[n \to \infty]{} 1. \qedhere
	\end{equation*}
\end{proof}

\bibliographystyle{elsarticle-harv} 
\bibliography{../../../Literature/literature.bib}

\end{document}